\documentclass[times, twoside]{zHenriquesLab-StyleBioRxiv}
\usepackage{blindtext}

\leadauthor{Benjamin} 

\usepackage{siunitx}
\usepackage{parskip}
\usepackage{placeins}
\def\N{\mathbb{N}}
\def\R{\mathbb{R}}

\DeclareMathOperator*{\argmax}{argmax}

\DeclareMathOperator*{\domain}{dom}
\def\partialto{\rightharpoonup}

\def\methodname{TopACT}

\def\genespace{\mathcal{G}} 
\def\gene{g} 
\def\genedim{D} 

\def\metricspace{X} 
\def\metricpoint{x} 

\def\spotgrid{\mathcal{S}} 
\def\spot{s} 

\def\scspace{\mathcal{C}} 
\def\singlecell{c} 

\def\celltypes{\mathcal{T}} 
\def\celltype{t} 
\def\celltyperv{T} 

\def\celltypemap{\tau} 
\def\sccelltypemap{\tau^{\mathrm{sc}}} 

\def\numcelltypes{K}

\def\expression{v} 
\def\exprv{V} 

\def\pooledexpr{v} 
\def\normpooledexpr{z} 

\def\scexpression{v^{\mathrm{sc}}} 

\def\normalizedspace{\mathcal{Z}} 
\def\normalizedvec{z} 
\def\normexprv{Z} 

\def\normalizedscexpression{z^{\mathrm{sc}}} 

\def\localclassifier{f}
\def\numspotsrv{Q}

\def\msca{\mathfrak{M}}

\def\topactmap{\mathfrak{T}}

\def\confidence{\theta} 
\def\rmax{r_\mathrm{max}} 
\def\rset{R}

\def\persmod{W} 
\def\persmap{\iota} 

\def\topspace{X} 

\def\metricspacemph{X}
\def\metricpointmph{x}

\def\sccountmatrix{C} 

\begin{document}

\title{Multiscale topology classifies and quantifies cell types in subcellular spatial transcriptomics}
\shorttitle{Multiscale topology classifies and quantifies cell types in subcellular spatial transcriptomics}

\makeatletter
\renewcommand\AB@affilsepx{; \protect\Affilfont}
\makeatother

\author[a]{Katherine Benjamin}
\author[b,c]{Aneesha Bhandari} 
\author[d,e]{Zhouchun Shang}
\author[d,e]{Yanan Xing}
\author[d]{Yanru An}
\author[f]{Nannan Zhang}
\author[d]{Yong Hou}
\author[a,g]{Ulrike Tillmann}
\author[b,c,1]{Katherine R.\ Bull}
\author[a,b,1]{Heather A.\ Harrington}

\affil[a]{Mathematical Institute, University of Oxford, Oxford OX2 6GG, United Kingdom}
\affil[b]{Wellcome Centre for Human Genetics, University of Oxford, Oxford OX3 7BN, United Kingdom}
\affil[c]{Nuffield Department of Medicine, University of Oxford, Oxford OX3 7FZ, United Kingdom}
\affil[d]{Beijing Genomics Institute-Shenzhen, Shenzhen, 518083, China}
\affil[e]{College of Life Sciences, University of Chinese Academy of Sciences, Beijing, 100049, China}
\affil[f]{Beijing Genomics Institute-Qingdao, Qingdao, 266555, China}
\affil[g]{Isaac Newton Institute for Mathematical Sciences, University of Cambridge, Cambridge CB3 0EH, United Kingdom}

\maketitle

\begin{abstract}
Spatial transcriptomics has the potential to transform our understanding of RNA expression in tissues. Classical array-based technologies produce multiple-cell-scale measurements requiring deconvolution to recover single cell information. However, rapid advances in subcellular measurement of RNA expression at whole-transcriptome depth necessitate a fundamentally different approach.
To integrate single-cell RNA-seq data with nanoscale spatial transcriptomics, we present a topological method for automatic cell type identification (TopACT). Unlike popular decomposition approaches to multicellular resolution data, TopACT is able to pinpoint the spatial locations of individual sparsely dispersed cells without prior knowledge of cell boundaries. Pairing TopACT with multiparameter persistent homology landscapes predicts immune cells forming a peripheral ring structure within kidney glomeruli in a murine model of lupus nephritis, which we experimentally validate with immunofluorescent imaging. The proposed topological data analysis unifies multiple biological scales, from subcellular gene expression to multicellular tissue organization. 
\end {abstract}


\begin{Affilfont}{{\textsuperscript{1}To whom correspondence should be addressed. \\ E-mail: katherine.bull@ndm.ox.ac.uk, harrington@maths.ox.ac.uk}}\end{Affilfont}

\section*{Introduction}
An open problem in spatial transcriptomics is the inference of information at the level of single cells \cite{marx_method_2021}. While recent experimental technologies enable whole-transcriptome measurement of gene expression at subcellular spatial resolutions, new computational methods are still required to extract single-cell information from these data. Here we provide mathematical tools to fill this gap and infer single-cell information from subcellular spatial measurements without prior knowledge of cell boundaries.

Spatial transcriptomics experiments have hitherto featured a trade-off between spatial resolution, transcriptome depth, and sample size~\cite{moses2022Museum}. Different experimental technologies have focused on different aspects. smFISH-based methods such as seqFISH~\cite{lubeckSinglecellSituRNA2014}, MERFISH~\cite{chenSpatiallyResolvedHighly2015}, and CosMX \cite{He2021.11.03.467020} enable subcellular transcript localization with high detection efficiency, but are difficult to scale up to the whole transcriptome on large samples. On the other hand, array-based methods such as ST/Visium~\cite{stahlVisualizationAnalysisGene2016} and Slide-Seq(v2)~\cite{rodriquesSlideseqScalableTechnology2019, stickels2021Highly} achieve whole-transcriptome depth but with resolution (\SIrange{10}{100}{\um}) above the diameter of a typical mammalian cell.

A wide range of methods have been developed to automatically predict cell types from multicellular resolution data~\cite{chen2022comprehensive, moses2022Museum} (Figure~1A, `Decomposition'). The majority of these methods focus on integration with single-cell or single-nucleus RNA-seq (sc/snRNA-seq) (Figure~1B), and can broadly be separated into two categories: imputation and decomposition~\cite{moses2022Museum}. Imputation methods, including Seurat~\cite{stuartComprehensiveIntegrationSingleCell2019} and Tangram~\cite{biancalani2021Deep}, impute the spatial coordinates of single-cell readings using the low-resolution spatial data as `lampposts.' Decomposition methods, typically based on non-negative matrix factorization~\cite{rodriquesSlideseqScalableTechnology2019, elosua-bayes2021SPOTlight} or statistical models~\cite{andersson2020Singlecell,  cable2021Robust, kleshchevnikov2022Cell2location, lopez2022DestVI, danaher2022Advances}, decompose multicellular readings to predict cell-type proportions. 

\begin{figure*}
\centering
\includegraphics[width=17.8cm]{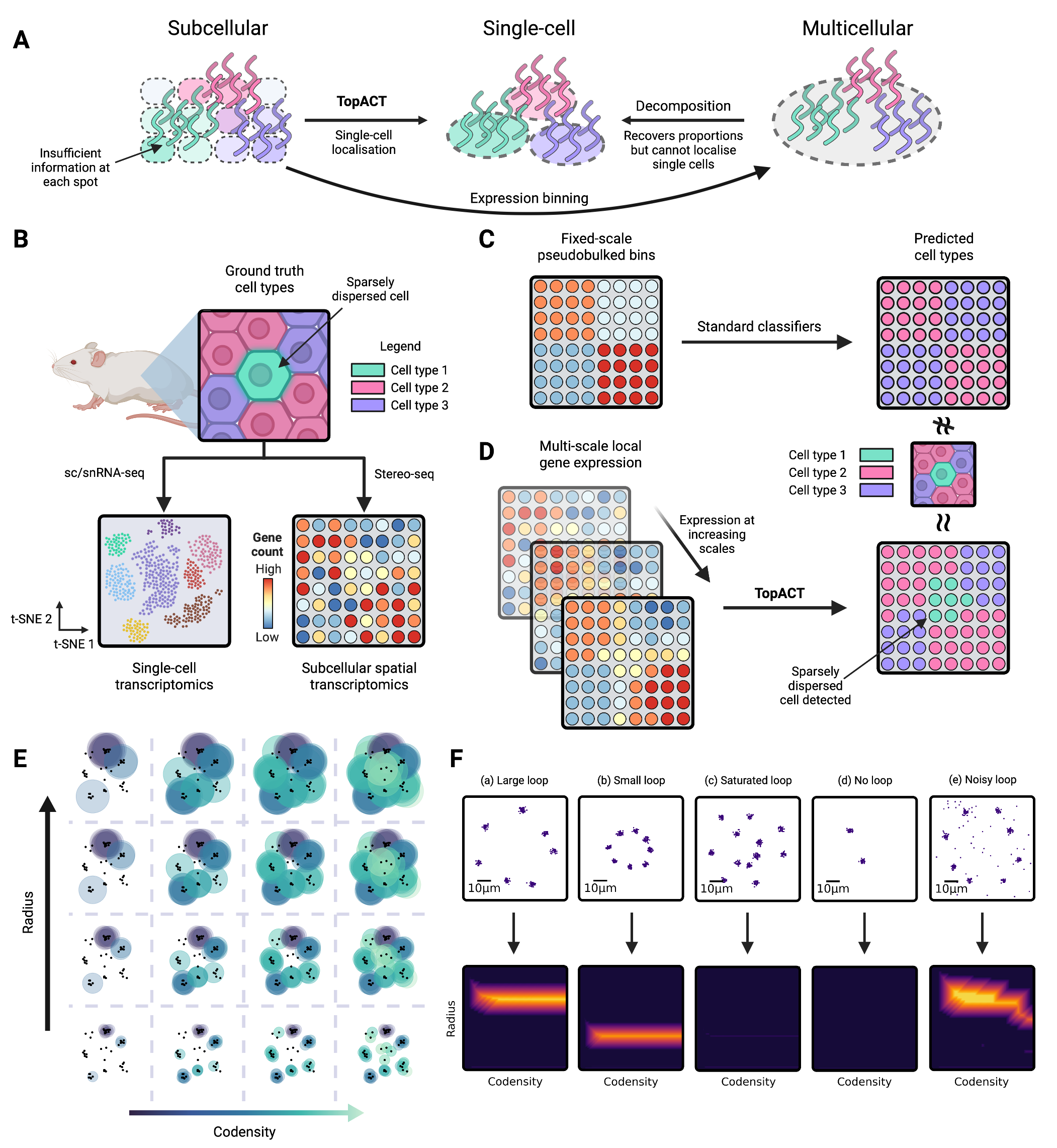}
\caption{\textbf{(A)} Spatial transcriptomics measurements can be grouped according to their relative scale: subcellular, single-cell, and multicellular. Existing methods decompose multicellular readings into cell type proportions. With subcellular data, it is necessary to travel in the opposite direction to reach the single-cell level. One approach is to aggregate expression in a fixed window to move to the multicellular regime and then decompose, however this fails to resolve individual cell loci. TopACT aggregates locally to move directly from the subcellular regime to the single-cell regime. \textbf{(B)} Automatic spatial cell type identification requires integration of single-cell and spatial transcriptomics. Here we use Stereo-seq, which produces gene expression readings on spots with center-to-center distance of 715nm. \textbf{(C)} Standard classifiers run on fixed-window expression bins can lose information about sparsely-dispersed cells caught between bin boundaries. \textbf{(D)} TopACT minimizes information loss by taking a flexible topological approach. Each spot is classified independently using local neighborhoods at multiple scales, accommodating heterogeneous cell sizes and varying per-spot transcriptional abundance. This flexibility allows TopACT to detect finer structural information at the single-spot level, including individual sparsely dispersed cells.
\textbf{(E)} An example radius-codensity bifiltration defined on a 2D point cloud. At each radius-codensity value, a ball of that radius is drawn on top of all points with at most the given codensity (i.e.\ sparseness). The hue of the ball indicates the codensity of the underlying point. The radius parameter therefore changes the scale of interaction between points, and the codensity parameter controls the level of noise reduction. Note that the loop structure is only detected at certain radius-codensity parameter values. 
\textbf{(F)} Archetypal spot assignment patterns (top row) and their corresponding MPH landscapes ($\lambda_1$ shown) (bottom row). (a) The large loop structure activates the landscape at high radius values. (b) The small loop structure activates the landscape at low radius values. (c) A saturated loop structure with clusters in the center of the loop does not activate the landscape. (d) A point cloud with no underlying loop structure does not activate the landscape. (e) The codensity parameter ensures that the landscape is still activated even in the presence of outliers and misclassifications.
}
\label{fig:overview}
\end{figure*}

    The inception of subcellular-resolution array-based technologies such as HDST ($\sim$\SI{2}{\um})~\cite{vickovic2019Highdefinition}, Seq-Scope ($\sim$\SIrange{0.5}{0.8}{\um})~\cite{cho2021Microscopic},
    and Stereo-seq (\SI{0.5}{\um})~\cite{chen2022Spatiotemporal} necessitates the development of bespoke mathematical methods for cell type identification. 
    Since the high spatial resolution of subcellular technologies is associated with low per-spot transcriptional abundance,
    precisely the opposite approach to decomposition is required. That is, to recover single-cell information from subcellular data, reads from neighboring spots must be aggregated to the single-cell level.
    When cell boundaries are unknown, a `fixed-window' approach is often taken, where a traditional decomposition method is used on a coarsely binned grid~\cite{chen2022Spatiotemporal, cho2021Microscopic} (Figure~1A, `Expression binning'). However, by its nature this approach discards the advantages of the high-resolution platform \cite{moses2022Museum}, rendering the inference of single-cell information impossible. Moreover, the fixed size and offset of the grid leads to underdetection of sparsely dispersed cells, such as immune cells, which are often of critical importance in clinical settings (Figure~1C).

Mathematically, the problem of automatic cell type identification in transcriptomics is to extract meaningful signal from extremely sparse, high-dimensional feature vectors. In the single-cell setting (where each cell in a sample is assigned a vector corresponding to the distribution of all of its RNA transcripts) standard clustering techniques identify cell types well. In the subcellular spatial setting, however, the transcripts in each cell are distributed among hundreds of spatial spots, meaning that each spot is assigned a sparse expression vector representing only a partial fragment of the total expression of its corresponding cell. The key challenge here is to intelligently aggregate these sparse vectors without \textit{a priori} knowledge of cell boundaries. This challenge is furthered by varying cell sizes, cell morphology and per-spot transcriptional abundance, motivating the development of a multiscale approach.

Here, we introduce a method for topological automatic cell type identification (TopACT) (Figure~1D) and combine this with multiparameter persistent homology (MPH)~\cite{carlsson2009Theory} to quantify the multiscale spatial organization of cell types. TopACT independently classifies the cell type of each spot using a local neighborhood, and dynamically chooses the size of the neighborhood based on the amount of information available around the spot. The end result is a spot-level cell type annotation which can be analyzed directly or used to determine the locations of individual cells. We build this information into a topological pipeline based on MPH landscapes~\cite{vipond2020Multiparameter} which enables quantification and comparison of the spatial distributions of individually resolved cells across multiple samples (Figures~1E and 1F).

As a proof of principle, we use random Voronoi diagrams to simulate cell spatial organization and run TopACT on corresponding synthetic subcelluar spatial transcriptomics data imputed from an snRNA-seq reference. We demonstrate that TopACT is able to produce spot-level cell type annotations with high accuracy, recovering single-cell-level structure that is inaccessible from fixed-window approaches. Next, we showcase TopACT on mouse kidney data generated by Stereo-seq~\cite{chen2022Spatiotemporal}. 
Mice treated with a toll-like receptor 7 (TLR7) agonist develop autoimmunity with kidney pathology modeling that observed in systemic lupus erythematosus (SLE), including mild renal immune infiltration~\cite{https://doi.org/10.1002/art.38298}, allowing us to test the ability of TopACT to detect pathological changes.
We show that the proposed method is able to spatially resolve individual immune cells.
MPH analysis then generates a hypothesis on immune cell organization, which we confirm with immunofluorescent~(IF) imaging.

\section*{Topological Model}

The subcellular spatial transcriptomics data considered in this work consist of a spatial grid of \emph{spots}, each of which is equipped with a \emph{gene expression} vector taken from $\mathbb{R}^D$ where $D \approx 25000$ is the number of genes in the genome. Each spot has an associated cell type, and we assume that the expression vectors are sampled from a random process dependent on this cell type. We are interested in the inverse problem of determining the cell type from the gene expression data. The challenge with subcellular data is that these readings are very sparse, with each expression vector typically containing only a handful of nonzero entries, rendering it impossible to determine the cell type of each spot in isolation. It is therefore necessary to aggregate the expression in neighboring spots to determine the corresponding cell types.

\subsection*{TopACT: Topological Automatic Cell Types} TopACT provides spot-level cell type annotations of subcellular spatial transcriptomics data. We assume that the aggregation of expression vectors from a sufficient number of spots of the same cell type yields an expression vector comparable to an sc/snRNA-seq expression vector of that cell type. We then use this assumption to approach spatial cell type classification as a transfer learning problem, leveraging the high efficacy of automatic cell type classification in the single-cell setting \cite{xie2021Automatic}. By default, TopACT learns from an annotated sc/snRNA-seq reference sample to construct a single-cell classifier. This classifier is applied at multiple scales of aggregation, producing a multiscale cell type confidence matrix at each spot. By analyzing this matrix, TopACT outputs a predicted cell type for each spot. See \textit{Materials and Methods} for further details.

\subsection*{MPH landscapes} We use MPH to quantify the spatial organization of cell types produced by TopACT. The input to MPH is a point cloud (here, the collection of all spots at which TopACT predicts a given cell type). MPH then constructs a radius-codensity \emph{bifiltration}: a collection of shapes parameterized by a choice of scale (radius) and level of noise reduction (codensity) (see Figure~1E). The output of MPH is a summary, called an \emph{MPH landscape} \cite{vipond2020Multiparameter}, of certain topological information contained in the bifiltration (here, the presence of loops). The landscape is a 2D image whose intensity at a given pair of radius and codensity parameters records the strength of the topological signal in the bifiltration at those parameters. Different loop structures yield different activation patterns in the MPH landscape, allowing for the inference of qualitative structural information (Figure~1F). In particular, the codensity parameter provides a robustness to misclassifications (Figure~1F~(d)). MPH landscapes can be leveraged to perform statistical analysis and machine learning, as has been demonstrated on immune cell patterns in tumors~\cite{vipond2021Multiparameter}. See \textit{Materials and Methods} for further details.

\section*{Description of data sets}

\subsection*{Synthetic Voronoi model}

\begin{figure}[!tbh]
\includegraphics[width=8.7cm]{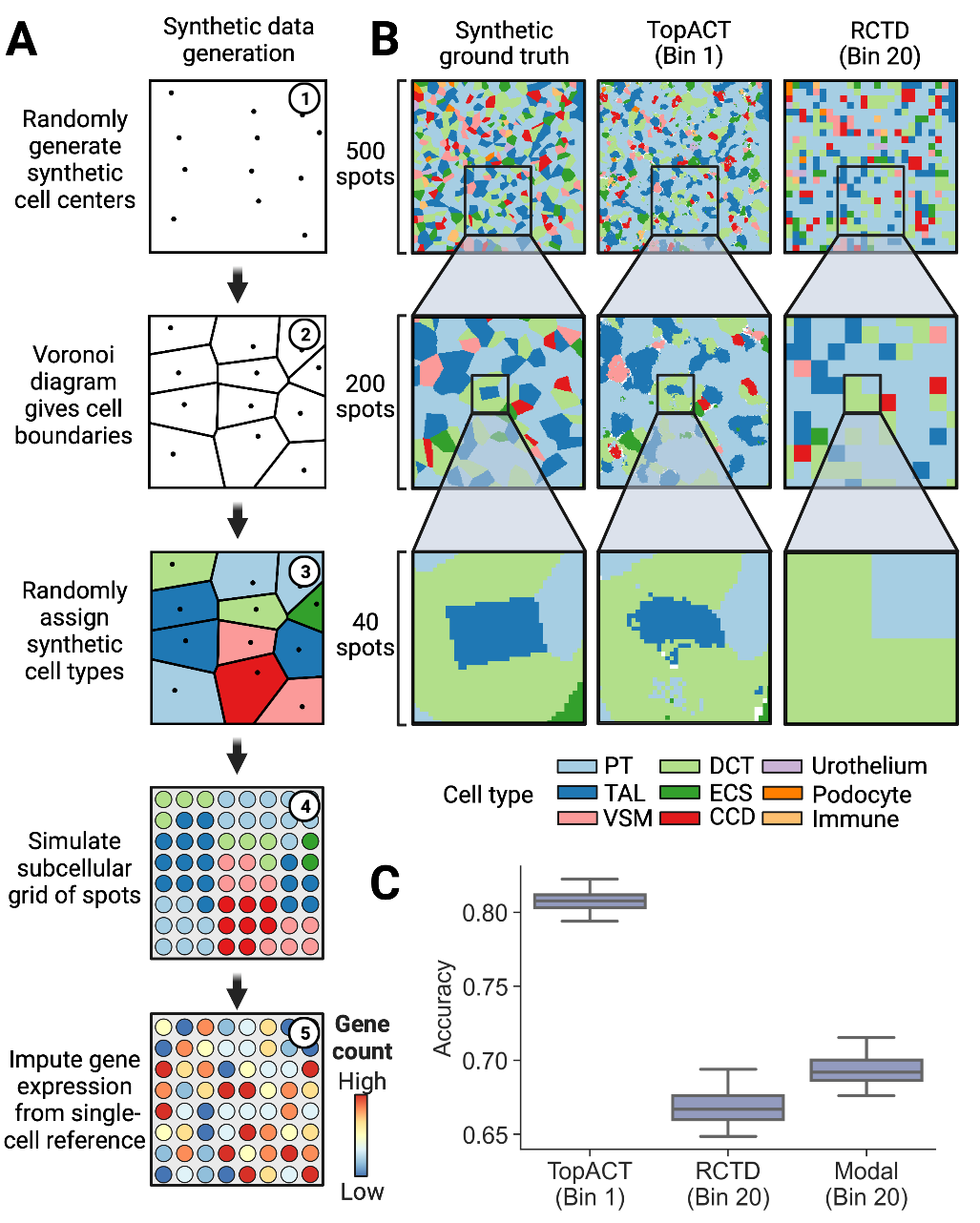}
\caption{Benchmarking TopACT with synthetic data. \textbf{(A)} A five-step process for generating synthetic subcellular spatial transcriptomics data with associated ground truth cell types. For full details, see \textit{Materials and Methods}. (1) Cell centers are generated uniformly at random from a unit square. (2) A Voronoi diagram is drawn to trace boundaries between cells. (3) Cell types are assigned to each region at random, in proportion to those in the reference kidney snRNA-seq data set. (4) A subcellular grid of spots is overlaid and annotated with cell types. (5) Synthetic gene counts are assigned to each spot using a Poisson process with parameters taken from the reference snRNA-seq data. \textbf{(B)} Sample output of various automatic cell type identification algorithms on synthetic data (see \textit{Materials and Methods}). From left to right: Ground truth, TopACT, RCTD \cite{cable2021Robust} at Bin 20. On the third row, note that TopACT is able to localize sparsely dispersed cells distributed in between the boundaries of fixed-window bins (\textit{c.f.}~Figure~1C-D). \textbf{(C)} Box plots of per-iteration accuracy of cell type classification methods on synthetic data (100 iterations). Accuracy of an iteration refers to the proportion of spots in that iteration to which the correct cell type is assigned. Underlying cell type maps change with each iteration. Modal refers to the theoretically-optimal fixed-window classification which assigns to each bin its most common ground truth cell type.}
\label{fig:synthetic}
\end{figure}

We use a two stage process to generate synthetic subcellular transcriptomics data with accompanying ground truth cell types (Figure~2A). We sample cell centers uniformly at random and then draw Voronoi diagrams to infer synthetic cell boundaries, inspired by work in digital pathology \cite{phillips2021Highly, schurch2020Coordinated}. We then use a Poisson process (with parameters inferred from snRNA-seq reference data, see \textit{Materials and Methods}) to impute gene expression counts on a subcellular grid overlaid on the diagram. The existence of ground truth cell types for these synthetic data allows for systematic benchmarking of cell type classification performance.

\subsection*{Mouse kidney model} Snap frozen murine kidneys were obtained from mice treated with topical Imiquimod, a TLR7 agonist, to both ears for 8 weeks, and control treated with petroleum jelly. Treated mice develop lupus-like renal disease with glomerular endocapillary proliferation, which includes proliferation of circulating immune cells that have migrated to the capillary tuft. \SI{10}{\um} cryosections from three mice (four slices from a control sample, and two and four slices respectively from two treated samples) were processed for spatial transcriptomics with Stereo-seq~\cite{chen2022Spatiotemporal}. Kidney tissue from the above samples was dissociated to single nuclei, partitioned and sequenced to generate snRNA-seq data~\cite{zheng_tenx}, yielding a matched single-nucleus and subcellular spatial transcriptomics data set. We cluster cells in the snRNA-seq data set using Seurat~\cite{hao2021seurat} and annotate cell types according to top marker genes. The spatial data for each sample consists of gene expression readings measured on a grid of \SI{220}{\nm} DNA nanoball spots with center-to-center inter-spot distance \SI{715}{\nm}. This data is represented by a \mbox{$D$-dimensional} expression vector ($D \approx 25000$) at each spot. See \textit{Materials and Methods} for further details.

\section*{Results}

We demonstrate TopACT on two data sets. We first benchmark the proposed method against the fixed-window decomposition approach on the synthetic Voronoi model. We then showcase TopACT on the mouse kidney model by pinpointing the location of sparsely dispersed immune cells. We demonstrate a significant increase in glomerular immune activity in treated samples, consistent with lupus-like immune infiltration. MPH then quantifies the spatial patterning of immune cells in treated kidney samples, predicting a ring structure in treated glomerular immune cells. This purely spatial-transcriptomic prediction is verified with IF imaging for leucocyte common antigen (CD45), showing increased glomerular immune staining in treated kidney, with a peripheral distribution pattern.

\subsection*{Benchmarking TopACT using synthetic data}

We use TopACT to classify cell types in benchmark data. TopACT is able to resolve fine structural detail including sparsely dispersed cells (Figure~2B). For comparison we run RCTD \cite{cable2021Robust}, an established decomposition method, using the standard fixed-window approach. We benchmark the accuracy of these methods, and find that TopACT surpasses both RCTD and the theoretically optimal fixed-window classification achieved by assigning each bin its modal ground truth cell type (Figure~2C). See \textit{Materials and Methods} for further details.

\subsection*{TopACT pinpoints immune cells in mouse kidney tissue} Inflammation is a key driver in multiple kidney diseases, including glomerulonephritis and diabetic nephropathy, but typically in these disorders immune cells do not aggregate or form large lymphoid structures within the kidney. Spatially resolving sparsely dispersed immune cells in kidneys is challenging; immune cells are small and metabolically inactive compared to abundant tubular cells, and therefore in a fixed bin containing multiple cell types tubular cell signatures will dominate.

Seeking to capture this behavior spatially, we apply a version of TopACT trained to detect immune and podocyte cells to Stereo-seq data from the mouse kidney model (Figure~3A). The proposed method is able to resolve the precise locations of individual podocyte (see Supplementary Note~2B) and immune cells (Figure~3B) in space. Comparing immune cell counts in glomerular vs non-glomerular regions of tissue (Figure~3C) identifies a statistically significant increase in glomerular immune activity in treated samples ($p = \num{8.3e-05}$) (Figure~3D), consistent with lupus-like immune infiltration.

\begin{figure}[!t]
\centering
\includegraphics[width=8.7cm]{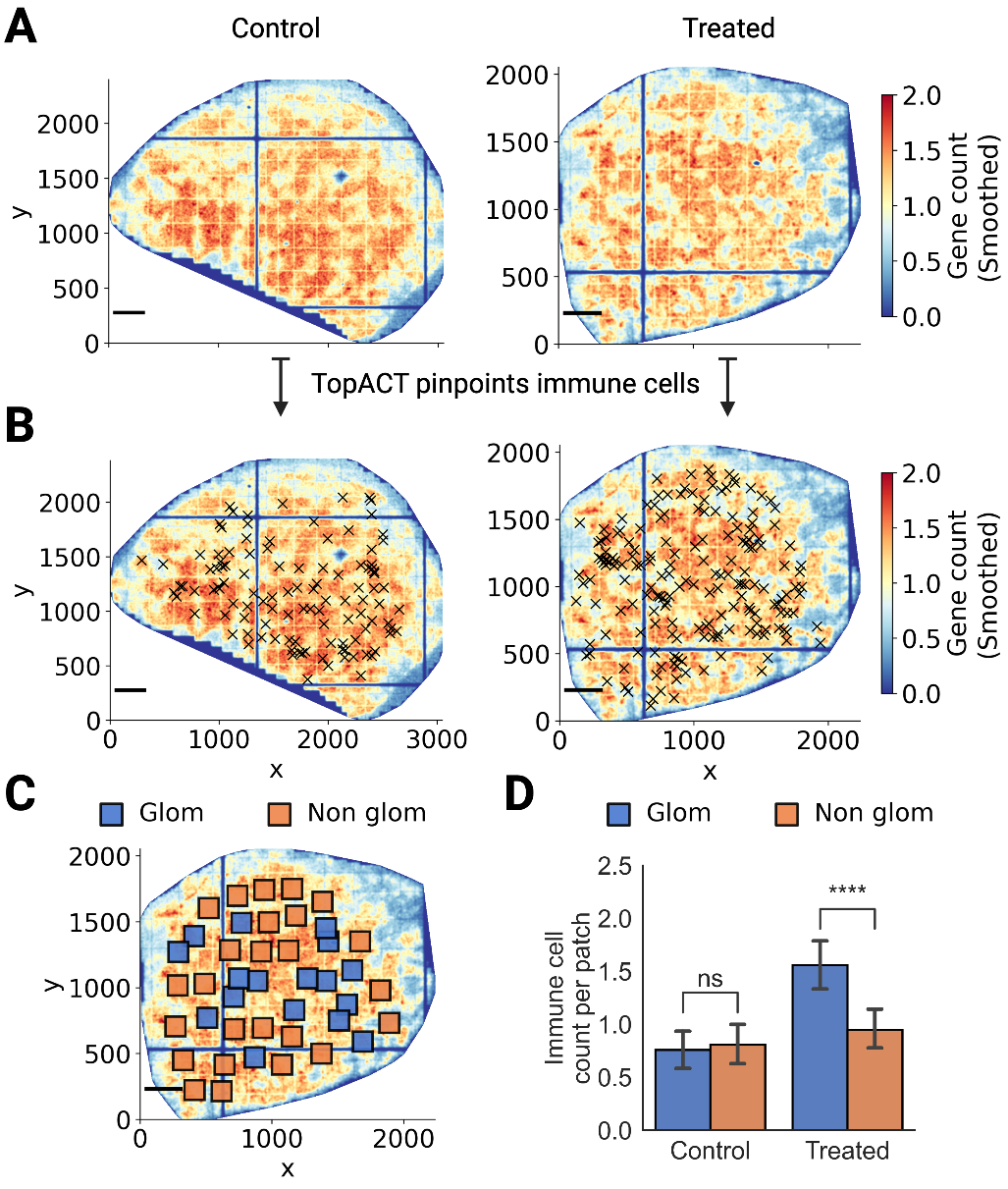}
\caption{TopACT pinpoints immune cells in mouse kidney tissue. \textbf{(A)}~Gene count density across example kidney sections. Axes show spot $x,y$ coordinates. Left: Control. Right: Treated. \textbf{(B)}~TopACT predicted immune cell loci (black 'x'). By default, the proposed method outputs an annotation of `immune' or `not immune' for each spot. Image analysis techniques (see SI~Appendix, Section~2B) are then used to extract single cell loci. Scale bars: 0.2mm. \textbf{(C)}~Example distribution of sampled square patches across a treated sample. Each patch is 150 spots ($\sim$107\textmu{}m) wide, slightly wider than the diameter of a typical mouse glomerulus. Glomerular patches (`Glom', blue square) are centered on glomeruli detected from Bin 20 data (see SI~Appendix, Section~2B). Non-glomerular patches (`Non glom', orange square) are randomly-sampled non-overlapping regions saturating the remaining tissue area (see SI Appendix, Section~2B). In total, we consider 269 glomerular patches (108 control, 161 treated) and 310 non-glomerular patches (130 control, 180 treated). Scale bars: 0.2mm. \textbf{(D)}~Mean immune cell count per patch, separated by treatment class and patch type. A significant increase in immune cell levels is observed in glomerular vs non-glomerular patches in treated samples ($p = 8.3 \times 10^{-5}$). This is consistent with a lupus-like model of immune infiltration. Error bars denote 95\% confidence intervals; stars indicate Welch's $t$-test $p$-values.}
\label{fig:immunecells}
\end{figure}

\subsection*{MPH characterizes immune cell spatial organization}
Kidney architecture is highly complex, and while existing spatial methods can place large-scale changes such disease driven fibrotic signatures in tissue context~\cite{Abedini2022.10.24.513598} and infer colocalization of renal immune cells, fine-scale mapping of infiltrating immune cells is hampered by both resolution and low detection of immune cells in spatial data~\cite{10.1172/jci.insight.147703}. The ability of TopACT to spatially pinpoint immune cells now enables the systematic quantification of their spatial organization.

MPH landscapes \cite{vipond2020Multiparameter} have proven highly successful in the study of the spatial organization of immune cells in tumors \cite{vipond2021Multiparameter}. Inspired by this work, we compute MPH landscapes from TopACT's output on small patches encapsulating each glomerulus in each sample (see Supplementary Note~2B). By taking the mean landscape across each sample, we arrive at an average summary of the multiscale immune cell topology around each glomerulus (Figure~4A(iv)). We observe that the average MPH landscape corresponding to treated kidneys is activated at high radius parameters, indicating the presence of large loops of immune cells. This leads to the hypothesis that this behavior is caused by the presence of a peripheral ring structure in immune cells infiltrating glomeruli in treated kidneys. We emphasize that this hypothesis is driven entirely by systematic analysis of spatial transcriptomics data.
\begin{figure*}
\includegraphics[width=17.8cm]{Figures/MPH}
\caption{Generating and validating a hypothesis on the spatial organization of immune cells using MPH.
\textbf{(A)} MPH analysis of glomerular immune cell distributions. Left: Control. Right: Treated. (i) Example distributions of glomerular patches (black squares, see Supplementary Note~2B). The background heatmap shows transcript density. Scale bars: 0.2mm. (ii) TopACT spot-level immune cell annotations on example patches. Each pixel represents a single Stereo-seq spot. Dark pixels indicate spots labeled `Immune' by TopACT. (iii) MPH landscapes ($\lambda_1$ shown here) for both example patches. The treated landscape is activated due to a loop structure in the underlying immune cell predictions. (iv) The average MPH landscapes for control and treated samples. The average landscape for control samples has low activation, consistent with a lack of infiltrating immune cells (\textit{c.f.}~Figure 1F (d)). The average landscape for treated samples is activated at high radius values, indicating the presence of large peripheral loop structures in infiltrating immune cells (\emph{c.f.}~Figure~1F (a) and (e).).
\textbf{(B)} Representative Immunofluorescence (IF) images of control and treated renal cortex. CD45 in magenta, nuclei stained with DAPI in blue. Glomeruli circled in white.
\textbf{(C)} Analysis of IF images. Left panel: mean fluorescent intensity in glomerular (`Glom') vs non-glomerular (`Non glom') regions in control and treated kidneys. The significant difference in treated samples ($p=5.1 \times 10^{-3}$) is consistent with TopACT predictions (\emph{c.f.}~Figure~3D). Right panel: ratio of mean intensity across the whole glomerulus to central glomerular region. Intensities measured across 151 glomeruli and ratio measured for 30 glomeruli per animal, across 3 control and 3 treated mice. The increased ratio in treated mice ($p=8.2 \times 10^{-4}$) is consistent with the MPH prediction of a peripheral ring structure in infiltrating immune cells. Error bars denote 95\% confidence intervals; stars indicate Welch's $t$-test $p$-values.}
\label{fig:mph}
\end{figure*}
To validate the predicted immune distributions at glomeruli, we performed IF imaging for CD45 in Imiquimod treated and control kidney sections (Figure~4B). Treated kidney showed more CD45 staining in glomeruli (Figure~4C, left panel), consistent with the TopACT prediction (Figure~3D). Moreover, quantification of global versus central CD45 intensity showed a significantly increased ratio in treated samples (Figure~4D, right panel), consistent with the MPH landscapes prediction of a peripheral ring structure.
\section*{Discussion}
In this work, we have introduced, implemented and applied TopACT, a multiscale method for topological automatic cell type classification. The proposed approach resolves cell type information at subcellular resolution, and zeros in on the location of elusive sparsely dispersed cells. By replacing the fixed-window view with a flexible, multiscale lens, TopACT achieves significantly higher accuracy in subcellular spatial cell type identification than the naive fixed-window approach.

We showcased the strengths of this approach on mouse kidney data generated by Stereo-seq \cite{chen2022Spatiotemporal}, which offers nanoscale-resolution, whole-transcriptome measurement of gene expression. We leveraged the high-dimensional, fine-scale detail of Stereo-seq to locate individual immune cells. By integrating TopACT with MPH landscapes \cite{vipond2020Multiparameter, vipond2021Multiparameter}, we revealed the spatial arrangement of glomerular immune cells in lupus nephritis, demonstrating the effectiveness of the topological approach in quantifying and elucidating tissue organization.

TopACT is a highly general and flexible mathematical methodology, offering a powerful approach to the problem of expression aggregation in the absence of cell boundaries. In future, the proposed method can be applied directly to higher-dimensional data including spatio-temporal and 3D data, as experimental technologies evolve in this direction.
\begin{small}
\section*{Materials and Methods}

\subsection*{Code and data availability}

The TopACT package is available at \url{https://gitlab.com/kfbenjamin/TopACT}. Code for synthetic data and experiments is available at \url{https://github.com/katherine-benjamin/topact-paper}. Clinical data and corresponding code will be made available on publication.

\subsection*{TopACT} TopACT operates on subcellular resolution spatial transcriptomics data. These data take the form of a grid of spots such that each spot in the grid has an associated expression vector in $\mathbb{R}^D$. The output of our method is a cell type annotation for each spot in the grid.

First, we use an annotated sc/snRNA-seq reference data set to construct an automatic single cell classifier, called a \emph{local classifier}. This classifier could in principle use any supervised learning approach, for example a neural network or random forest, but a simple and effective choice is the Support Vector Machine \cite{boser1992Training, cortes1995Supportvector, scikit-learn}. The only restriction is that the classifier must take as input an expression vector and output a probability vector over all cell types. We achieve this by using Platt scaling \cite{platt1999Probabilistic}.

Given a local classifier, TopACT then proceeds as follows. Fix a spot $s$ in the grid. For a given radius $r \geq 0$, consider a ball of radius $r$ drawn about $s$. Let $X_r$ be the sum of all the expression vectors of spots within this ball. Feeding $X_r$ as input to the local classifier produces a probability vector $v_r$ over the cell types. For a list $r_1 \leq r_2 \leq \dots \leq r_k$ of radii, these vectors can be combined into a multiscale cell type confidence matrix $A = [v_{r_1}, v_{r_2}, \dots, v_{r_k}]$. Now, pick a confidence threshold $\theta \in (0,1).$ For a spot $s$, let $j$ be minimal such that the most likely cell type $i$ at scale $r_j$ has confidence $A_{ij} \geq \theta$. We set the cell type of $s$ to be $i$ if such a $j$ exists. In words, the cell type assigned to the spot $s$ is the cell type predicted by the local classifier at the smallest possible scale at which a confident prediction can be made. For a full technical description of the TopACT method, see Supplementary Note~1A.

\subsection*{Synthetic data generation} We use a two-stage process to generate synthetic benchmark data, first generating a synthetic cell type map and then imputing gene expression. Firstly, a synthetic grid of spots with cell type annotations is produced. We sample 625 points uniformly at random from the unit square $[0,1] \times [0,1]$, taking these points to be cell centers. We draw a Voronoi diagram (computed using the implementation in SciPy \cite{scipy} based on Qhull \cite{qhull}) based on these points to simulate cell boundaries. Cell types are then assigned at random to each Voronoi region, in proportion to the cell type abundances in the snRNA-seq data. These cell types are then applied to a $500 \times 500$ grid of spots overlaid on the unit square. The end result is a grid of spots, each annotated with a cell type. Next, we impute the gene expression at each spot using a Poisson process with parameters inferred from the mouse kidney snRNA-seq data described below. This process is based on a simplified version of the model described in \cite{cable2021Robust}. In detail, for a cell type $T$ and gene $g$ let $\lambda_{Tg}$ denote the mean expression of gene $g$ over all cells in the snRNA-seq data set with cell type $T$. If a spot $s$ is assigned the cell type $T$, we then model the expression $v_{sg}$ of gene $g$ at $s$ by
\begin{equation}
    v_{sg} \sim \mathrm{Poisson}(\alpha \lambda_{Tg})
\end{equation}
where $\alpha=\exp(-7.3)$ is a fixed parameter determining the transcriptional abundance. To model zero-inflation, we then select \SI{20}{\percent} of spots uniformly at random to be assigned zero reads, regardless of the Poisson-modeled expression.

\subsection*{Synthetic data analysis} We ran TopACT directly on synthetic data, with an SVM local classifier trained on the same snRNA-seq reference data set used for generation (see Supplementary Note~2B). For fixed-window Bin 20 analysis, we split the $500 \times 500$ synthetic grid into square bins, each covering a $20 \times 20$ region of spots. Bin 20 was chosen so that each bin matches the mean area of a synthetic cell. Moreover, at Bin 20 the resulting grid approximates \SI{10}{\um} resolution, which is considered the ``sweet spot'' for single-cell analysis \cite{marx_method_2021}. We then summed the expression over all spots in each region. RCTD \cite{cable2021Robust} was run on couplet mode with default settings, using the same snRNA-seq reference data set, and we assigned each bin the RCTD predicted `first type'. For the modal cell type classification, we assigned to each bin its most frequent ground truth cell type.

\subsection*{MPH Landscapes}
MPH tracks how the topological features (here, loops) of a shape evolve as certain parameters are varied. Given an input point cloud, we record the first persistent homology ($H_1$) of its associated Rips-codensity bifiltration. This information is summarized in a sequence, called an \emph{MPH landscape}~\cite{vipond2020Multiparameter}, of functions $\lambda_k \colon \R \times \R \to \R$ for $k=1, 2, \dots$, Given a radius parameter $s$ and a codensity parameter $t$, the value $\lambda_k(s,t) \in \mathbb{R}$ roughly describes the significance of the $k$-th most significant topological feature in the bifiltration at those parameter values. Here, we focus on $\lambda_1 \colon \R \times \R \to \R$ which describes the significance of the most significant such feature. For a full introduction to MPH, see Supplementary Note~1B.

Here we computed average control and treated MPH landscapes for TopACT predicted immune cell points clouds (see also Supplementary Note~2B).
MPH was computed with RIVET (\url{https://github.com/rivetTDA/rivet/}) and converted to MPH landscapes using the code from \cite{vipond2021Multiparameter} (\url{https://github.com/MultiparameterTDAHistology/SpatialPatterningOfImmuneCells}).

\subsection*{Statistical tests} Statistical tests were performed using the Python packages Scipy (\url{https://www.scipy.org/}) and statannotations (\url{https://github.com/trevismd/statannotations}).

\subsection*{Animals} Female BALB/cOlaHsd mice were purchased from Envigo (Bicester, Oxford) at 5 weeks of age. Animals were housed in specific pathogen free individually ventilated cages under project licence P84582234, kept on a 12-hour light/dark cycle from 8:00-20:00, with food and water freely available. All experiments were carried out in compliance with UK Home Office Guidelines and the Animals Scientific Procedure Act 1986 (amended 2013) and reported in line with the ARRIVE guidelines. Mice were treated topically with either \SI{5}{\percent} Aldara (Imiquimod) cream (Meda Pharmaceuticals) or Vaseline (Unilever, Surrey, UK) control on both ears, 3 times weekly for 8 weeks.

\subsection*{Spatial RNA sequencing} Stereo-seq was performed at the Beijing Genomics Institute as previously described \cite{chen2022Spatiotemporal}. Capture chips were loaded with DNA nanoballs (DNB) generated by rolling circle amplification of random 25 base pair (\si{bp}) oligonucleotides. Single end sequencing (MGI DNBSEQ-Tx) was performed to determine the DNB co-ordinate identity at each spatial location on the chip, followed by ligation of \SI{22}{bp} polyT and \SI{10}{bp} molecular identity oligos to the DNB.  
\SI{10}{\um} Kidney tissue sections were cryosectioned from OCT embedded frozen blocks and adhered to the chip surface, fixed in methanol, stained with nucleic acid dye (Thermo Fisher Scientific, Q10212) for imaging, and incubated at \SI{37}{\degreeCelsius} with \SI{0.1}{\percent} pepsin (Sigma, P7000) for \SI{12}{\min} to permeabilize. After permeabilization, reverse transcription and cDNA amplification were performed and the Agilent 2100 was used to check the range of cDNA fragments. cDNA was interrupted by in-house Tn5 transposase and amplified, and the fragments double-selected. After screening, libraries were subjected to Agilent 2100 quality inspection. Finally, the double-selection libraries were constructed into libraries suitable for the MGI DNBSEQ-Tx sequencing platform through circularization steps and were sequenced to collect data (\SI{50}{bp} for read 1 and \SI{100}{bp} for read 2).

\subsection*{Single-nucleus RNA sequencing} Tissue from the same frozen sections used for spatial transcriptomics was used to perform complementary single nuclei RNA seq. Single nuclei were isolated as previously described with minor modifications \cite{pmid30586455}. Briefly, kidney tissues were placed into a \SI{2}{\ml} Dounce homogenizer (Sigma) with \SI{2}{\ml} pre-chilled Homogenization Buffer (\SI{10}{mM} Tris pH 8.0 (Thermo Fisher), \SI{250}{mM} sucrose (Sigma), \SI{1}{\percent} BSA (Sangon Biotech), \SI{5}{mM} MgCl2 (Thermo Fisher), \SI{25}{mM} KCl (Thermo Fisher), \SI{0.1}{mM} DTT (Thermo Fisher), 1X Protease inhibitor cocktail (Roche), \SI{0.4}{U\per\ul} RNase inhibitor (MGI), \SI{0.1}{\percent} NP40 (Roche)). After incubation on ice for \SI{10}{\min}, tissues were homogenized by 10 strokes of the loose pestle A and filtered with \SI{70}{\um} cell strainer (Falcon). The homogenate was further homogenized with 10 strokes by tight pestle B, filtered using \SI{30}{\um} cell strainer (Sysmex) into \SI{15}{\ml} conical tube, and centrifuged at \SI{500}{\gram} for \SI{5}{\min} at \SI{4}{\degreeCelsius}. The pellet was resuspended in \SI{1}{\ml} blocking buffer (1X PBS (Thermo Fisher), \SI{1}{\percent} BSA, \SI{0.2}{U\per\ul} RNase inhibitor) and centrifuged at \SI{500}{\gram} for \SI{5}{\min}, this step was repeated once. The pellet was resuspended using cell resuspension buffer (MGI) at concentration of \SI{1000}{nuclei\per\ul} for further library preparation.
snRNA-seq libraries were prepared using DNBelab C Series Single-Cell Library Prep Set (MGI, \#1000021082) \cite{Liu818450}. Droplets were generated from a single nuclei suspension, followed by emulsion breakage, bead collection, reverse transcription, and cDNA amplification to generate barcoded libraries. Indexed libraries were constructed following the manufacture’s protocol, quantified using Qubit ssDNA Assay Kit (Thermo Fisher Scientific, Q10212) and sequenced using DNBSEQ-T1 at the China National GeneBank (Shenzhen, China) with read length \SI{41}{bp} for read1, \SI{100}{bp} for read2, and \SI{10}{bp} for sample index. 

\subsection*{Single-nucleus clustering}
snRNA-Seq data were analyzed using Seurat \cite{hao2021seurat}. Nuclei were filtered on gene count $<500$ or $> 3500$, and mitochondrial $\si{\percent} > 5$. Data was log normalized, variable features identified, and linear transformation scaling performed. The first 30 principle components were selected and clusters identified using the `FindClusters' method in Seurat with a resolution of $0.6$. The `FindAllMarkers' function was used to identify genes that characterized each cluster and differential expression of genes was tested between clusters. Cluster annotation was performed manually based on the top markers, applying in-house expertise in renal physiology with reference to the literature. 

\subsection*{Immunofluorescence} Kidneys were harvested, snap frozen and embedded in OCT embedding matrix (Fisher Scientific, Loughborough, UK). \SI{10}{\um} tissue sections on SuperFrost Plus glass slides (VWR, Lutterworth, UK) were fixed in two changes of ice-cold Acetone (Sigma), washed with TBS and blocked for \SI{1}{hr} at room temperature (RT) with blocking buffer. Slides were incubated overnight at \SI{4}{\degreeCelsius} with anti-CD45 (clone: 30-F11, eBioscience, Thermofisher, Paisley, UK), followed by incubation with Alexa Fluor 594 conjugated goat anti-mouse IgG (H+L) (Invitrogen) secondary antibody, for \SI{1}{hr} at RT and counterstained with DAPI for \SI{10}{\min} at RT. Slides were mounted with VECTASHIELD\textregistered{} Antifade Mounting Media (Vector Laboratories, Peterbourough, UK) and imaged using a Leica DMi8 Ca+ Imager at the Wellcome Cellular Imaging Core using fixed laser intensities. TIFF images were analyzed using ImageJ (FIJI) software. Mean intensity was measured across glomerular regions, manually defined based on bright field and nuclei, and non-glomerular regions comprising the whole field of view with the glomerular regions subtracted. Ratios were calculated per glomerulus using mean intensity for whole glomerulus (global) and a central glomerular region.

\begin{acknowledgements}
We thank John Todd and Richard Cornall for introductions, and Joshua Bull for helpful discussions. We thank the support provided by China National GeneBank. K.B.\ is supported by the J.T.\ Hamilton and EPSRC Scholarship. K.B., H.A.H., and U.T.\ are grateful for the support provided by the UK Centre for Topological Data Analysis EPSRC grant EP/R018472/1. H.A.H.\ gratefully acknowledges funding from the Royal Society RGF\textbackslash{}EA\textbackslash{}201074 and UF150238. K.R.B.\ is supported by the Medical Research Council and Kidney Research UK, grant MR/R007748/1.
For the purpose of Open Access, the authors have applied a CC BY public copyright licence to any Author Accepted Manuscript (AAM) version arising from this submission. Figures created with BioRender.com.\end{acknowledgements}

\begin{contributions}
K.B., U.T., K.R.B., and H.A.H.\ designed the research.
A.B., Z.S., Y.X., Y.A., N.Z., Y.H., and K.R.B.\ contributed data.
K.B., U.T., K.R.B, and H.A.H.\ contributed new analytic tools.
K.B., A.B., and K.R.B.\ analyzed data.
All authors wrote the paper.
\end{contributions}

\begin{interests}
The chip, procedure, and applications of Stereo-seq are covered in pending patents. Z.S., Y.X., Y.A., N.Z., and Y.H.\ are employees of BGI and have stock holdings in BGI.
\end{interests}
\end{small}

\onecolumn
\newpage

\captionsetup*{format=largeformat}

\section{Analysis techniques}

Here we describe in detail the topological methods used in the main text. We start by introducing TopACT, a method for topological automatic cell type identification on subcellular spatial transcriptomics data. We then describe multiparameter persistent homology (MPH) landscapes, a method from Topological Data Analysis which we use to detect topological features in TopACT predicted immune cell distributions.

\subsection{Cell type classification for subcellular spatial transcriptomics}

We begin by describing mathematical model for subcellular spatial transcriptomics data. We then detail how TopACT can be applied on such data to extract cell type classifications at the spot level.

\subsubsection{Subcellular spatial transcriptomics model}

	We begin by abstracting the notion of a spatial transcriptomics experiment. A general (non-spatial) transcriptomics experiment can be thought of as a collection of objects (for example, in single-cell transcriptomics the objects are cells), each equipped with a cell type $\celltype$ and an expression vector $\expression \in \R^\genedim.$ The vector $\expression$ measures the number of reads in each of the $\genedim{}$ genes in the genome $\genespace$, and it is assumed that these are sampled from random variables corresponding to the cell type $\celltype$. The key difference in a spatial transcriptomics experiment is that the objects are now equipped with a notion of distance, giving rise to a metric space. The present section formalizes this notion.
	\paragraph{Experimental setup}
	We begin with the following preliminary objects:
	\begin{enumerate}
		\item A metric space $\metricspace$ called a \emph{sample};
	    \item A finite subset $\spotgrid \subset \metricspace$ of \emph{spots};
	    \item A finite ordered set $\genespace = \{\gene_1, \dots, \gene_\genedim\}$ of $\genedim$ \emph{genes};
        \item A finite ordered set $\celltypes = \{\celltype_1, \dots, \celltype_\numcelltypes\}$ of $\numcelltypes$ \emph{cell types}.
	\end{enumerate}
	These items together can be seen as a mathematical abstraction of a typical spatial transcriptomics experimental setup: we aim to measure the expression of each gene in \(\genespace\) across the sample \(\metricspace\), by taking readings from each spot in \(\spotgrid\). These readings are determined by the underlying cell type in $\celltypes$ associated to each spot. 
	
	In this setting, an experimental reading can be thought of as an assignment of an expression $v_{\spot g} \in \R$ for each spot $\spot \in \spotgrid$ and gene $\gene \in \genespace.$ Equivalently, making use of the ordering on $\genespace,$ we have a map \begin{equation}\expression \colon \spotgrid \to \R^\genedim,\end{equation} where $\expression(\spot)_i = \expression_{\spot \gene_i}$ for each $\spot \in \spotgrid$ and $1 \leq i \leq \genedim$.
	
	\paragraph{Expression model} \label{sec:expressionmodel}
	We now describe our model for how these expression assignments arise in practice. Underlying each experiment, we assume there is a set $\celltypes$ of disjoint \emph{cell types}, and that for each cell type \(\celltype \in \celltypes\) and gene \(\gene \in \genespace\) there is a corresponding random variable \(\exprv_{\celltype \gene}\) giving the count of the gene $\gene$ measured in a cell of type $\celltype$.
	
	A subcellular spatial transcriptomics experiment can be seen as a partial assignment
	\begin{equation}
	    \celltypemap \colon \metricspace \partialto \celltypes
	\end{equation}
	of a cell type to some of the points in \(\metricspace\). Given such an assignment, for each point \(\metricpoint \in \domain  \celltypemap\) we model \begin{equation}
	    \expression_{\metricpoint \gene} \sim \exprv_{\celltypemap(\metricpoint)\gene},
	\end{equation} and we can assign \(\expression_{\metricpoint \gene}=0\) whenever \(\metricpoint \notin \domain \celltypemap\). Restricting these values to spots in $\spotgrid$, we recover an experimental reading.
	
	We emphasize that in this subcellular model, each spot is assigned \emph{at most} one cell type. In the case of multicellular spatial transcriptomics, as seen with data produced by e.g.\ ST/Visium~\cite{stahlVisualizationAnalysisGene20161} and Slide-Seq(v2)~\cite{rodriquesSlideseqScalableTechnology20191, stickels2021Highly1}, this assumption will not hold, as each spot records transcripts from multiple distinct cells.
	
\subsubsection{TopACT method description}

    Given that the expression vector $\expression(\metricpoint) \in \R^\genedim$ assigned to a point $\metricpoint \in \metricspace$ depends on its cell type $\celltypemap(\metricpoint)$, a natural objective is to deduce the cell type map $\celltypemap$ given the expression map  $\expression.$ In the case of single cell transcriptomics, where each expression vector contains sufficient information to deduce a cell type, this is a relatively straightforward task. In contrast, subcellular spatial data typically suffer from very low read counts, and it is therefore necessary to aggregate readings from neighboring spots in order to recover enough information to reliably predict a cell type. However, it is not clear how best to perform this aggregation without prior knowledge of cell boundaries.
    
    Our approach is to assume that there exists a local neighborhood around each spot that belongs entirely to a single cell type. By combining the expression readings from this neighborhood, one obtains a pseudo-single-cell reading that can be classified by existing techniques. This yields a classification for each individual spot. The challenge now is to identify the correct scale at which to draw the neighborhood, and we resolve this by taking a `multiscale' approach.
    
	\paragraph{Local classifier definition}

	Let $\celltyperv$ be a $\celltypes$-valued random variable and $\numspotsrv$ a positive-integer-valued random variable. We are going to study a random variable describing the aggregated gene expression of $\numspotsrv$ spots that all assigned the cell type $\celltyperv$.
	
	For any cell type $\celltype \in \celltypes$ let \begin{equation}
	    \exprv_\celltype = (\exprv_{\celltype, \gene_1}, \dots, \exprv_{\celltype, \gene_\genedim} )
	\end{equation} 
	be the $\R^\genedim{}$-valued random variable describing the total expression over all genes of the cell type $\celltype.$ Then, if $\exprv_\celltype^1, \dots, \exprv_\celltype^K$ are i.i.d\ copies of $\exprv_\celltype$, set $\Sigma \exprv_\celltype =  \sum_{i=1}^\numspotsrv \exprv_\celltype^i$ and
	\begin{equation}
	    \normexprv = {\Sigma\exprv_{\celltype}}/{\lVert \Sigma\exprv_{\celltype} \rVert_1}.
	\end{equation}
$\normexprv$ is therefore the normalized sum of $\numspotsrv$ expression readings independently drawn from the cell type $\celltype$.

We say that a \emph{local classifier} is any method that estimates the probability of each cell type given an observed normalized expression reading. More specifically, recalling that the cell types have an ordering $\celltypes = \{\celltype_1, \dots, \celltype_\numcelltypes\}$, we say that a local classifier is a function \begin{equation}\localclassifier \colon \normalizedspace \to [0,1]^\numcelltypes,\end{equation} where $\normalizedspace = \{ \normalizedvec \in [0,1]^\genedim : \lVert \normalizedvec \rVert_1 = 1 \}$, such that
\begin{equation}\label{eq:localclassifier}
    \localclassifier(\normalizedvec)_i \approx \mathbb{P}\left(\celltyperv = \celltype_i \mid \normexprv = \normalizedvec \right)
\end{equation}
for all $1 \leq i \leq \numcelltypes.$
    \paragraph{Producing a local classifier from sc/snRNA-seq data}
    
    We can use single-cell or single-nucleus reference data sets to estimate the effect of the different cell types on the gene expression behavior and produce a local classifier. In detail, we take a collection \(\scspace\) of single cell samples along with a gene expression map
    \[ \scexpression \colon \scspace \to \R^\genedim \]
        and a cell type map
    \[ \sccelltypemap \colon \scspace \to \celltypes. \]
    
    From this information, we seek a classifier that takes as input normalized expression vectors and outputs probability distributions over the cell types in \(\celltypes\). To do this, we normalize each expression vector:
    \begin{equation}
        \normalizedscexpression(\singlecell) = \frac{\scexpression(\singlecell)}{\lVert \scexpression(\singlecell)\rVert_1}.
    \end{equation}
    The input-output pairs \(\left(\normalizedscexpression(\singlecell), \sccelltypemap(\singlecell)\right)\) then form training data for any standard supervised learning platform. In our case, we use a linear support vector machine (SVM) \cite{boser1992Training1, cortes1995SupportVector1} and estimate probabilities with Platt scaling \cite{platt1999Probabilistic1} to produce a local classifier $\localclassifier$.
    
    \paragraph{Multiscale confidence matrix}
    
    Let $\localclassifier$ be a local classifier.
    In order to classify a point $\metricpoint \in \metricspace$ it may be necessary to aggregate expression readings around $\metricpoint.$ Write $B(x, r) = \{y \in X : d(x, y) \leq r\}$ for the closed ball of radius $r$ in $X$ centered on $x$. We define the aggregated gene expression
    \begin{equation}\label{eq:aggregate} \pooledexpr(\metricpoint, r) = \sum_{\spot \in B(\metricpoint,r) \cap \spotgrid} \expression(\spot) \in \R^\genedim,\end{equation}
    and, if this is non-zero, set
    \begin{equation}
        \normpooledexpr(\metricpoint,r) = \pooledexpr(\metricpoint,r) / \lVert \pooledexpr(\metricpoint,r) \rVert_1 \in \normalizedspace.
    \end{equation}
    In words, $\normpooledexpr(\metricpoint,r)$ describes the normalized gene expression about $\metricpoint$ at radius $r$. Then, if $\localclassifier$ is a local classifier as defined in \eqref{eq:localclassifier}, one obtains a probability vector \begin{equation}f(\normpooledexpr(\metricpoint, r))\end{equation} which can be interpreted as a cell type classification at the scale $r.$
    
    Given an ordered collection  $\rset = (r_1 \leq \dots \leq r_L)$ of radii one then obtains a sequence of corresponding probability vectors, which can be combined into an $L \times \numcelltypes$ matrix $\msca{}^\metricpoint$ defined by
    
    \begin{equation}
        \msca{}^\metricpoint_{ij} = f(\normpooledexpr(\metricpoint, r_i))_j
    \end{equation}
    which we call a \emph{multiscale confidence matrix.} Here $\msca^\metricpoint_{ij}$ records the confidence in cell type $t_j$ at scale $r_i$ around the point $\metricpoint.$
    
\paragraph{Extracting cell type annotations}

Given a multiscale confidence matrix $\msca^\metricpoint$, we would like to extract a cell type annotation for the spot $\metricpoint.$ The general principle followed by \methodname{} is that one should use the smallest scale possible to classify a point, because this minimizes the chance that the aggregated expression has been taken from surrounding cells of a different type.

Let $\confidence \in [0,1]$ be a \emph{confidence hyperparameter}. The \emph{classification index} $i_\confidence = i_\confidence(\metricpoint)$ of $\metricpoint$ is
\begin{equation}
    i_\confidence = \inf \left\{ i \in \{1, \dots, L\} : \lVert \msca_i^\metricpoint \rVert_1 \geq \confidence \right\}.
\end{equation}
In other words, $r_{i_\confidence(\metricpoint)}$ is the lowest scale at which a cell type was predicted with confidence at least $\confidence$ at the point $\metricpoint$.

We now define the \emph{TopACT predicted cell types} with respect to the collection $\rset$ and confidence threshold $\confidence$:
\begin{equation}
    \topactmap_{\rset, \confidence} \colon \metricspace \partialto \celltypes.
\end{equation}
If $i_\confidence(\metricpoint) < \infty$ then we set $\topactmap_{\rset, \confidence}(\metricpoint)$ to be the cell type with the highest predicted probability at scale $r_{i_\theta(\metricpoint)}$. Precisely, it is $\topactmap_{\rset, \confidence}(\metricpoint) = \celltype_j$ where $j$ maximizes the value of $\msca^\metricpoint_{i_\confidence(\metricpoint) j}$.\footnote{A tie between multiple cell types can be resolved by equipping the cell types with an order of precedence. Note that if $\theta > 0.5$ then a tie can never occur.} If $i_\confidence(\metricpoint)= \infty$, i.e.\ if no scale produced sufficient confidence, then we do not specify a cell type. In other words, we have that $\mathrm{dom}\left( \topactmap_{\rset, \confidence}\right) = \{ \metricpoint \in \metricspace : i_\confidence(\metricpoint) < \infty \}. $

\paragraph{Restricting TopACT to a square grid}
\label{sec:squaregrid}
In the experiments considered in the manuscript, we work with either simulated or real-world Stereo-seq \cite{chen2022Spatiotemporal1} data.
For Stereo-seq experiments, we assume that spots are evenly spaced on a 2D square lattice. More precisely, we assume that the metric space $\metricspace$ is a subspace of $\R^2$ and the set of spots is \begin{equation} \label{eq:squaregrid} \spotgrid = ([I] \times [J]) \cap \metricspace\end{equation} for some $I, J \in \N$, where $[k] = \{1, \dots, k\}$ for any $k \in \N$.

By further equipping $\R^2$, and therefore $X$, with the $\ell_\infty$ norm, it follows that the neighborhoods $B(\metricpoint, r)$ are squares in $X$. In particular, for a spot $\spot \in \spotgrid$ the critical values $r_0 \leq r_1 \leq \dots$ for which $B(\metricpoint, r_i) \cap \spotgrid$ changes are precisely $r_i = i \in \N$. In this setting, then, we set $\rset = (0, 1, 2, \dots, \rmax{})$ for some maximal radius parameter $\rmax \in \N$. Algorithm \ref{alg:topact} demonstrates how to produce \methodname{} cell type annotations from these assumptions.

\begin{algorithm}[tbh] \caption{\methodname{} (Square grid)} \label{alg:topact}
\hspace*{\algorithmicindent}\textbf{Input:} 
\begin{minipage}[t]{.8\textwidth}
\begin{itemize}[leftmargin=2.6mm, noitemsep]
 \item[] $M, N$: the dimensions of the spot grid;
 \item[] $V$: an $M \times N \times \genedim$ array where $V_{ijk}$ is the expression of gene $g_k$ at the spot $(i,j)$;
 \item[] $\localclassifier \colon \normalizedspace \to [0,1]^\numcelltypes$: a local classifier;
\item[] $\confidence$: a confidence hyperparameter;
\item[] $\rmax$: the maximum radius.
\end{itemize}
\vspace{-0.2mm}
\end{minipage}

\hspace*{\algorithmicindent}\textbf{Output:} The TopACT cell type assignment $\topactmap_{(0, \dots, \rmax), \confidence} \colon [M] \times [N] \partialto \celltypes$.
\begin{algorithmic}[1]
\State $\celltypemap{} \gets \emptyset$ \Comment{An empty cell type assignment}
\For{$s = (i,j) \in [M] \times [N]$} 
    \State $r \gets 0 \in \N$
    \State $\pooledexpr{} \gets 0 \in \R^\genedim$
    \While{$r \leq \rmax$ and $s \not\in \domain \celltypemap{}$}
        \For{all $s' = (i', j') \in [M] \times [N]$ such that $\lVert s - s' \rVert_\infty = r$} \Comment{Update pooled expression}
            \State $\pooledexpr \gets \pooledexpr + V_{i'j'}$
        \EndFor
        \If{$\pooledexpr \neq 0$}
            \State $\normalizedvec \gets \pooledexpr / \lVert \pooledexpr \rVert_1$\Comment{Normalize expression for input to local classifier}
            \State $k^* \gets \argmax_{1 \leq k \leq \numcelltypes} \localclassifier(\normalizedvec{})_k$
            \If{$\localclassifier(\celltype)_{k^*} \geq \confidence$}
                $\celltypemap(\spot) \gets \celltype_{k^*}$\Comment{Sufficient confidence to classify spot}
            \EndIf 
        \EndIf
        \State $r \gets r + 1$ \Comment{Increment radius to the next critical value}
    \EndWhile
\EndFor
\State \Return{$\celltypemap$}
\end{algorithmic}
\end{algorithm}

We remark that this setup may differ for different spatial transcriptomics technologies. For example, the spots may lie on a hexagonal grid as in HDST \cite{vickovic2019Highdefinition1} or be randomly distributed as in Seq-Scope \cite{cho2021Microscopic1}. Our method is general and applies equally to any such specification, including 3D or spatio-temporal data. TopACT only requires some notion of distance between the spots in $\spotgrid.$

\subsection{Multiparameter persistent homology}

Here, we provide a brief introduction to the theory of multiparameter persistent homology, a mathematical tool which tracks the presence of topological features in data as given parameter values are varied. In the case of a single parameter (where the parameter being varied is typically a type of scale), it is possible to represent this persistent homology via an interpretable summary known as a barcode. However, it is often the case that more than one parameter may need to be varied (here, to control for the presence of misclassifications), and in this case there is no complete, canonical description of the various topological information contained in the corresponding persistent homology. Nevertheless, there do exist interpretable summaries in this setting, and in this work we make use of the \emph{multiparameter persistence landscapes} of Vipond \cite{vipond2020Multiparameter1}. For a more complete theoretical discussion of multiparameter persistence see \cite{botnan2022introduction1}, and for a thorough exploration of applications to immune cell spatial distributions see \cite{vipond2021Multiparameter1}.

\subsubsection{Single parameter persistent homology}

We give a brief overview of single-parameter persistent homology in order to motivate our use of the multiparameter generalization. See e.g.~\cite{otter2017roadmap1} for a more complete introduction. The aim of persistent homology is to extract information about the multiscale topology of a point cloud in $\R^n$. This is achieved by assigning to such a point cloud a parameterized sequence of nested shapes, called a \emph{filtration}, and then studying how the topological features (connected components, holes, voids, and higher dimensional analogues) of these shapes vary as the parameter changes.

\begin{definition}\label{def:filtration}
    Let $X$ be a topological space. A \emph{single parameter filtration of $X$} is a collection of subspaces $X_t$ of $X$ for each $t \in \R$ such that $X_s \subset X_t$ whenever $s \leq t$.
\end{definition}
An instructive example of a filtration is given by drawing a ball of radius $t$ around each point in a point cloud, and letting these balls grow with the parameter $t$. We will refer to this as the \emph{\v{C}ech filtration}. It is intuitive that new topological features (e.g.\ loops) will appear (\emph{birth}) and be filled in (\emph{death}) as $t$ increases and the balls grow. To make this precise, we will make use of the mathematical theory of \emph{homology}\footnote{Not to be confused with the entirely distinct notion of homology in biology.}.

Homology (see \cite{allenhatcher2002Algebraic1} for a technical description) assigns to a given shape $X$ a vector space\footnote{Here we take homology with coefficients in $\mathbb{F}_2$} $H_k(X)$, a basis for which corresponds to the $k$-dimensional topological features in $X$. Furthermore, this assignment satisfies the \emph{functoriality} property, a consequence of which is that an inclusion $X \subset Y$ induces a linear map $H_k(X) \to H_k(Y)$ which encodes how the topological features of $X$ are realized in the larger space $Y$. The upshot of this is that one can associate to a filtration $(X_t)_{t \in \R}$ a sequence of vector spaces $(H_k(X_t))_{t \in \R}$ linked by linear maps $H_k(X_s) \to H_k(X_t)$ whenever $s < t$. This sequence contains precisely all the information of how the $k$-dimensional topological features of the filtration varies as the parameter changes.

Remarkably, the structure theorem of Zomorodian and Carlsson \cite{zomorodian2004computing1} guarantees that, under reasonable assumptions on the underlying filtration, all of this information can be completely summarized in a single object called a \emph{barcode}. This barcode is a multiset of intervals, one for each topological feature in the filtration, whose endpoints indicate the birth and death parameters of the underlying feature.

A key strength of persistent homology is that it is robust to perturbations: slightly moving the points in the point cloud does not significantly alter the resulting barcode. However, single-parameter persistent homology is highly sensitive to outliers. For example, introducing a single point into the center of a ring of points will kill the resulting $1$-dimensional feature, drastically altering the resulting barcode. This is problematic when there is a possibility of  misclassifications, as is the case in this work.

\subsubsection{Multiparameter persistence}

To deal with outliers, we will introduce a second parameter which filters out points below a certain density. We thus need to generalize the single-parameter filtration introduced in Definition \ref{def:filtration}. We first endow $\R^n$ with the following poset structure which defines when $\mathbf{s} \leq \mathbf{t}$ for pairs $\mathbf{s}, \mathbf{t} \in \R^n$:

\begin{equation}
    (s_1, \dots, s_n) \leq (t_1, \dots, t_n) \iff s_i \leq t_i \text{ for all $i \in \{1, \dots, n\}$.}
\end{equation}

\begin{definition}
    Let $\topspace$ be a topological space. An \emph{$n$-parameter filtration of $\topspace$} is a collection of subspaces $\topspace_\mathbf{t}$ of $\topspace$ for each $\mathbf{t} \in \R^n$ such that $\topspace_\mathbf{s} \subset \topspace_\mathbf{t}$ whenever $\mathbf{s} \leq \mathbf{t}.$
\end{definition}
Note that when $n=1$ we recover the single-parameter filtration of Definition \ref{def:filtration}.

A key example of a multiparameter filtration for $n=2$ extends the single-parameter \v{C}ech filtration to depend on an extra filtering function. We will take this filtering function to be \emph{codensity}.
\begin{definition}\label{def:codensity}
    Suppose $P \subset \R^n$ is a finite point cloud. For $k\geq 1$, the \emph{$k$-codensity} function $\rho_k \colon P \to \R$ is given by $$\rho_k(p) = \frac{1}{k}\sum_{i=1}^k \lVert p -  p_{(i)}\rVert$$ for each $p \in P$, where $p_{(i)}$ is the $i$-th nearest neighbor of $p$ in $P$.
\end{definition}
\begin{definition}
    Suppose $P \subset \R^n$ is a finite point cloud with corresponding codensity $\rho_k \colon P \to \R$. For a given subset $Q \subset P$ write $$B_r(Q) = \{\metricpointmph \in \metricspacemph : \lVert x - q \rVert \leq r \text{ for some $q \in Q$} \}$$ for the $r$-neighborhood about $Q$. For each $k \geq 1$, there is a $2$-parameter filtration given by
    $$\metricspacemph_{r, t} = B_r\left(\rho_k^{-1}\left((-\infty, t]\right)\right).$$ We call these filtrations \emph{\v{C}ech-codensity filtrations}.
\end{definition}
Note that fixing the codensity parameter $t$ and varying $r$ is equivalent to taking the single-parameter \v{C}ech filtration over the filtered point cloud $\rho_k^{-1}((-\infty, t])).$ The idea is that at low values of $t$ only the densest points are included, and so by increasing $t$ it is possible to track the topological effect of including less dense points into the filtration.
In practice, the \v{C}ech-codensity filtration is impractical to work with. In its place, it is typical to consider the \emph{Rips-codensity} filtration, which can be seen as an approximation to the \v{C}ech-codensity filtration.

Just as before, by applying homology one arrives at a collection of vector spaces $(H_k(X_\mathbf{t}) )_{\mathbf{t} \in \R^n}$ linked by linear maps describing how topological features are included from one set of parameter values to another. This is a special example of the following more general kind of algebraic object.

\begin{definition}
    An \emph{$n$-parameter multiparameter persistence module $\persmod$} consists of the following data:
    \begin{itemize}
        \item A vector space $\persmod_\mathbf{t}$ for each $\mathbf{t} \in \R^n$;
        \item A linear map $\persmap_{\mathbf{s}, \mathbf{t}} \colon \persmod_\mathbf{s} \to \persmod_\mathbf{t}$ for each pair $\mathbf{s}, \mathbf{t} \in \R^n$ whenever $\mathbf{s} \leq \mathbf{t}$.
    \end{itemize}
    In addition, the linear maps must satisfy:
    \begin{itemize}
        \item $\persmap_{\mathbf{t}, \mathbf{t}} = \mathbf{1}_{\persmod_\mathbf{t}}$ for each $\mathbf{t} \in \R^n$
        \item $\persmap_{\mathbf{s}, \mathbf{t}} \circ \persmap_{\mathbf{r}, \mathbf{s}} = \persmap_{\mathbf{r}, \mathbf{t}}$ for each triple $\mathbf{r}, \mathbf{s}, \mathbf{t} \in \R^n$ whenever $\mathbf{r} \leq \mathbf{s} \leq \mathbf{t}$.
    \end{itemize}
    In other words, a multiparameter persistence module is precisely a functor $\persmod \colon (\R^n, \leq) \to \mathrm{Vect}$.
\end{definition}

In the single-parameter setting, the decomposition of these persistence modules gives rise to the barcode. However, in the multi-parameter setting there is no such complete description of the information inside a persistence module. To arrive at an interpretable summary it is therefore necessary to define a representation that discards some of the included information.

\subsubsection{MPH landscapes} For single-parameter persistent homology, Bubenik proposed persistence landscapes as a vectorization of the barcode \cite{bubenik2015Statistical1}. Later, Vipond generalized this notion to give the \emph{multiparameter persistence landscape} \cite{vipond2020Multiparameter1}, which is a vectorized invariant of multiparameter persistence modules. Vipond et al.~\cite{vipond2021Multiparameter1} later applied this invariant to the study of immune cell spatial patterning in tumors, directly motivating the application of multiparameter persistence landscapes in this work.

\begin{definition}
    Let $\persmod$ be an $n$-parameter multiparameter persistence module. The \emph{(multiparameter) persistence landscape associated to $\persmod$} is a function $\lambda \colon \N \times \R^n \to \R$ given by
    $$\lambda(k, \mathbf{t}) = \sup \left\{ \varepsilon > 0 : \beta_{\mathbf{t - \varepsilon 1}, \mathbf{t + \varepsilon 1}} > k\right\},$$
    where $\beta_{\mathbf{s, t}} = \mathrm{rank}(\iota_{\mathbf{s,t}} \colon \persmod_\mathbf{s} \to \persmod_\mathbf{t})$ and we take the convention that the supremum of the empty set is $0$. We will also write $\lambda_k \colon \R^n \to \R$ for the function $\lambda_k(\mathbf{t}) = \lambda(k, \mathbf{t})$.
\end{definition}

If the persistence module is acquired by taking homology of a multiparameter filtration, one can think of $\lambda_k(\mathbf{t})$ as describing the `significance' of the $k$-th most significant topological feature in the filtration at the parameter $\mathbf{t} \in \R^n$. In particular, $\lambda_1$ records the significance of the most significant feature.

A particularly useful property of persistence landscapes is that they can be averaged, which is notably not a property of barcodes: there exist simple examples of two barcodes with multiple Fr\'echet means. Given a family $\lambda^1, \dots, \lambda^N$ of persistence landscapes, their \emph{average persistence landscape $\overline{\lambda}$} is taken to be the pointwise average:
\begin{equation}\label{eq:mpaverage}
    \overline{\lambda}_k(\mathbf{t}) = \frac{1}{N} \sum_{i=1}^N \lambda^i_k(\mathbf{t}).
\end{equation}

\section{Data analysis}

Here we describe in more detail the data analysis carried out in the main text. We begin by describing our implementation of the TopACT methodology for square grids. We then describe how TopACT and MPH landscapes are applied to clinical Stereo-seq data. Note that our analysis of synthetic Voronoi data is described entirely in the main text (\textit{Synthetic Voronoi model} and \textit{Benchmarking TopACT using synthetic data}).

\subsection{\methodname{} implementation}

We provide a Python package for \methodname{} with support for classification of 2D square grids as described in Algorithm \ref{alg:topact}. The package is modular and flexible, so it is possible for example to substitute in a custom local classifier in place of the provided SVM classifier.

\paragraph{Gene filtering}

We performed minimal gene filtering in our experiments, making use of the ability of SVM classifiers to maintain performance with high dimensionality. For each spatial sample, we restrict to the set of genes present in both the sample and the snRNA-seq reference. If the user wishes to make use of a different local classifier, such as a neural network, it is likely that more gene filtering would be necessary.

\paragraph{Learning a local classifier} \label{sec:svm}

For our experiments, we make use of the annotated snRNA-seq data described in the main text (\textit{Single-nucleus RNA sequencing} and \textit{Single-nucleus clustering}). We first filter out all genes that do not appear in the spatial data under consideration. Let $C$ be the resulting snRNA-seq count matrix, so that $\sccountmatrix_{ij}$ is the number of counts of gene $j$ in sample $i$. The rows of the matrix $C$ are normalized via the transformation
\begin{equation}
    \sccountmatrix'_{ij} = \log \left( \frac{10^5 \sccountmatrix_{ij}}{\sum_{j'}\sccountmatrix_{ij'} } +1\right).
\end{equation}
The columns of $\sccountmatrix'$ are then scaled to have unit variance, and the resulting feature matrix is used as training data for a linear SVM classifier. We use the SVM implementation in scikit-learn 1.1.1 \cite{scikit-learn1} with default settings. The same normalization pipeline is used when the local classifier is applied for spatial classifications.

\paragraph{Hyperparameters} \methodname{} requires two hyperparameters: a maximal radius $\rmax{}$ and a confidence threshold $\confidence$. We set the maximal radius $\rmax{}$ to be $9$, which for the mouse kidney data approximates the radius of a single cell (9 spots $\approx$ \SI{6.4}{\um}). We also set a minimum radius $r_1=3$ to improve efficiency. \methodname{} is implemented so that different values of the confidence threshold $\theta$ can be manually compared after classification is run, allowing for $\theta$ to be varied per experiment. We set $\theta = 0.5$ for synthetic data and $\theta = 0.7$ for mouse kidney data. Note that synthetic data features classifications from 9 classes compared to just 3 for clinical data, justifying the use of a lower confidence hyperparameter.

\subsection{Stereo-seq data analysis}

Here we describe how we ran TopACT on the Stereo-seq \cite{chen2022Spatiotemporal1} data described in the main text. We analyzed ten slices (4 from a control kidney, and 2 and 4 respectively from two treated kidneys). See the main text (\textit{Mouse kidney model}) for further description of this data set.

\begin{figure}
\centering
\includegraphics[width=15.77cm]{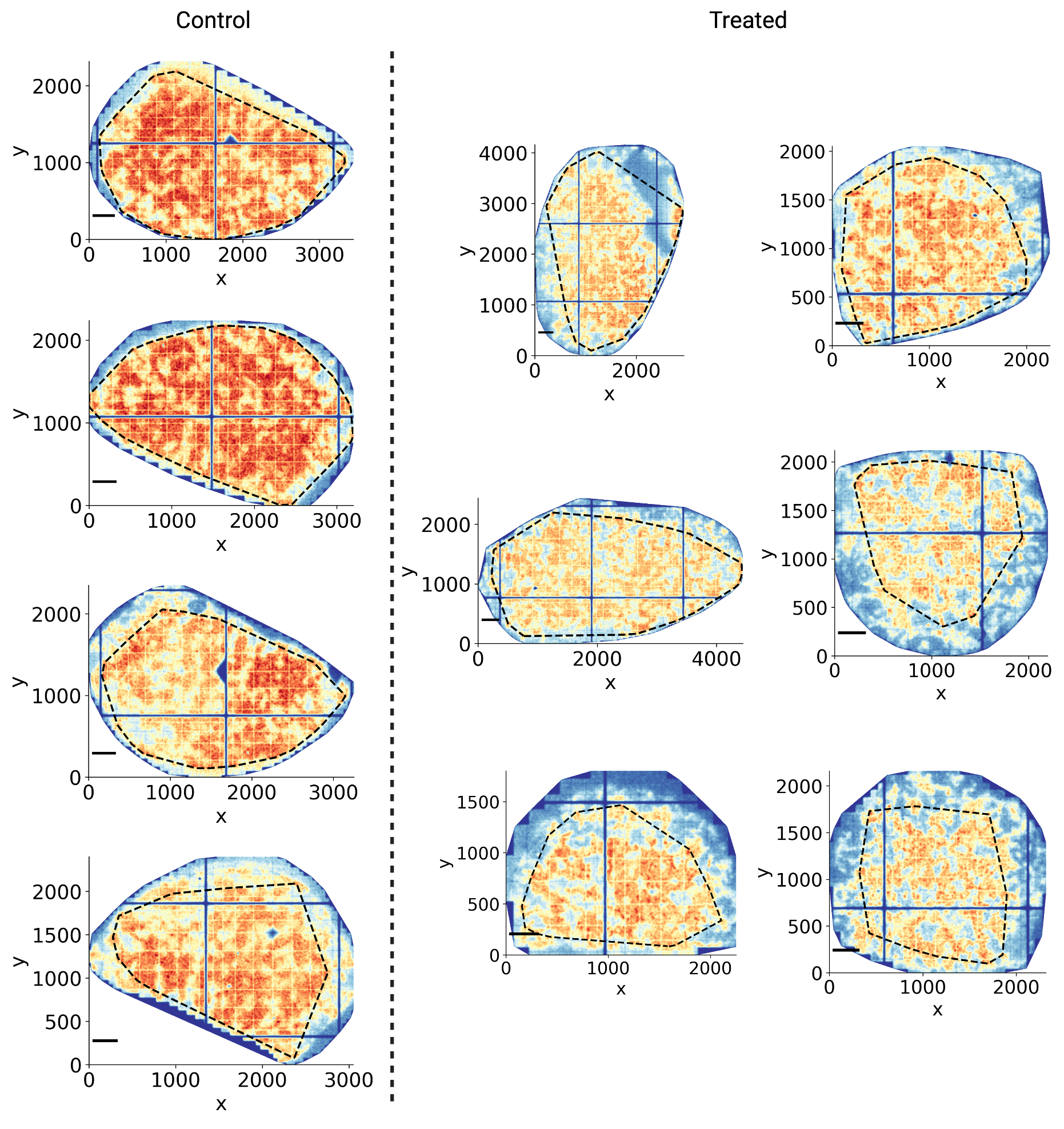}
\caption{Defining sample boundaries based on transcript density. Each axis shows a single sample (Left: Control. Right: Treated). Black dashed lines show convex hulls of high-density regions in each sample, which are used to restrict later analysis. Background heatmaps show smoothed transcript count. Scale bars = 0.2mm.}
\label{fig:masks}
\end{figure}

\paragraph{Defining sample boundaries}
For computational efficiency, we restricted cell type classifications to a convex hull approximating the underlying sample shape. In detail, we consider the gene density at each spot, i.e.\ the mean number of reads in a square of side length 21 centered on the spot. We then take the boundary of a sample to be the convex hull of all points with sufficiently high ($>5$) density (see Figure \ref{fig:masks}). This ensures that computation time is not wasted on the boundary region of the sample, which we found had insufficient transcript counts to yield meaningful classifications.

\paragraph{TopACT output} We ran TopACT on each of the ten mouse kidney samples, restricted to the previously described regions. For the local classifier, we used an SVM classifier trained from snRNA-seq data annotated with three classes:
\begin{enumerate}
    \item Podocyte cell,
    \item Immune cell,
    \item Other cell type.
\end{enumerate}

\begin{figure}
\centering
\includegraphics[width=\textwidth]{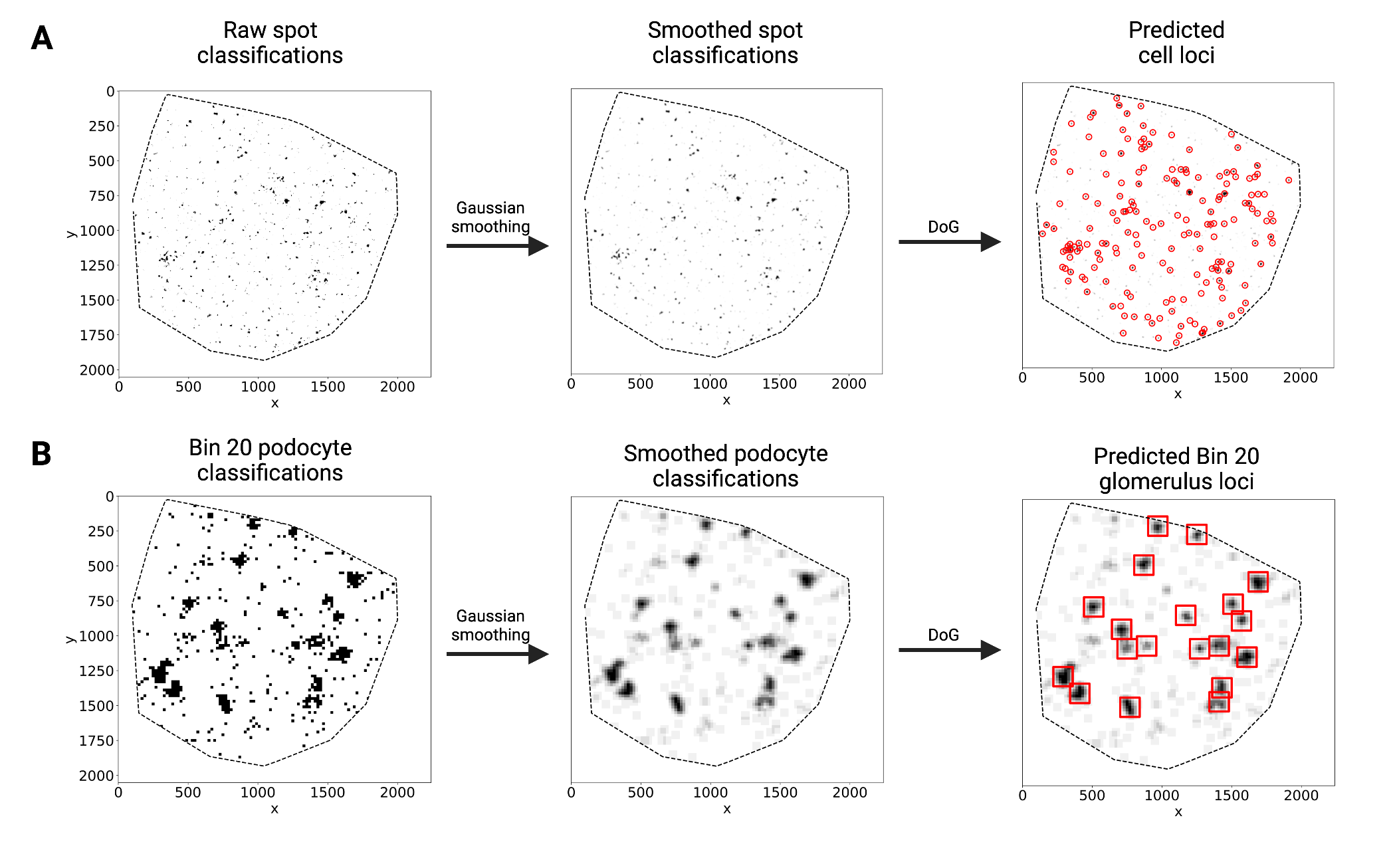}
\caption{\textbf{(A)} Extracting cell loci from spot-level cell type predictions. First, a binary image is produced indicating spots assigned the given cell type. Then, Gaussian smoothing is applied to produce a grayscale image. Finally, DoG blob detection~\cite{lowe2004Distinctive1, scikit-image1} is used to detect regions of high density of the given cell type. These regions are taken as predicted cell loci. In this example, immune cell loci are detected. \textbf{(B)} Extracting ground truth glomerulus loci from Bin 20 cell type predictions. The pipeline is identical to that set out in (A), but run specifically on podocyte predictions at Bin 20.}
\label{fig:cellprediction}
\end{figure}

\paragraph{Detecting single cells from \methodname{} output} \label{sec:dog} We use a standard image analysis pipeline to extract single cell loci from \methodname{} output. In detail, for a given cell type and sample we produce a binary image representing spots that are classified with the given cell type. We then perform a difference of Gaussians (DoG) blob detection \cite{lowe2004Distinctive1} (computed using scikit-image \cite{scikit-image1}) on a Gaussian smoothing of this binary image to extract single cell loci (see Figure \ref{fig:cellprediction} A).

\paragraph{Validation via podocyte prediction} \label{sec:glom}

To validate the performance of \methodname{} on real-world data, we tested its ability to detect podocyte cells. Podocyte cells colocalize almost exclusively with glomeruli, which are large enough that they can be easily detected by existing methods at Bin 20 (i.e.\ with expression pooled into square bins with a side-length of 20 spots). This provides an ideal ground truth for validation. In detail, we used Seurat~\cite{hao2021seurat1} with the same procedure as in the main text (\textit{Single-nucleus clustering}) to produce a cell type annotation of each mouse sample at Bin 20 resolution. A pipeline similar to that used for single-cell detection was then used at Bin 20 resolution to detect regions of high podocyte density (Figure \ref{fig:cellprediction} B). We took these regions to be ground truth glomeruli.

\begin{figure}
\centering
\includegraphics[width=15.77cm]{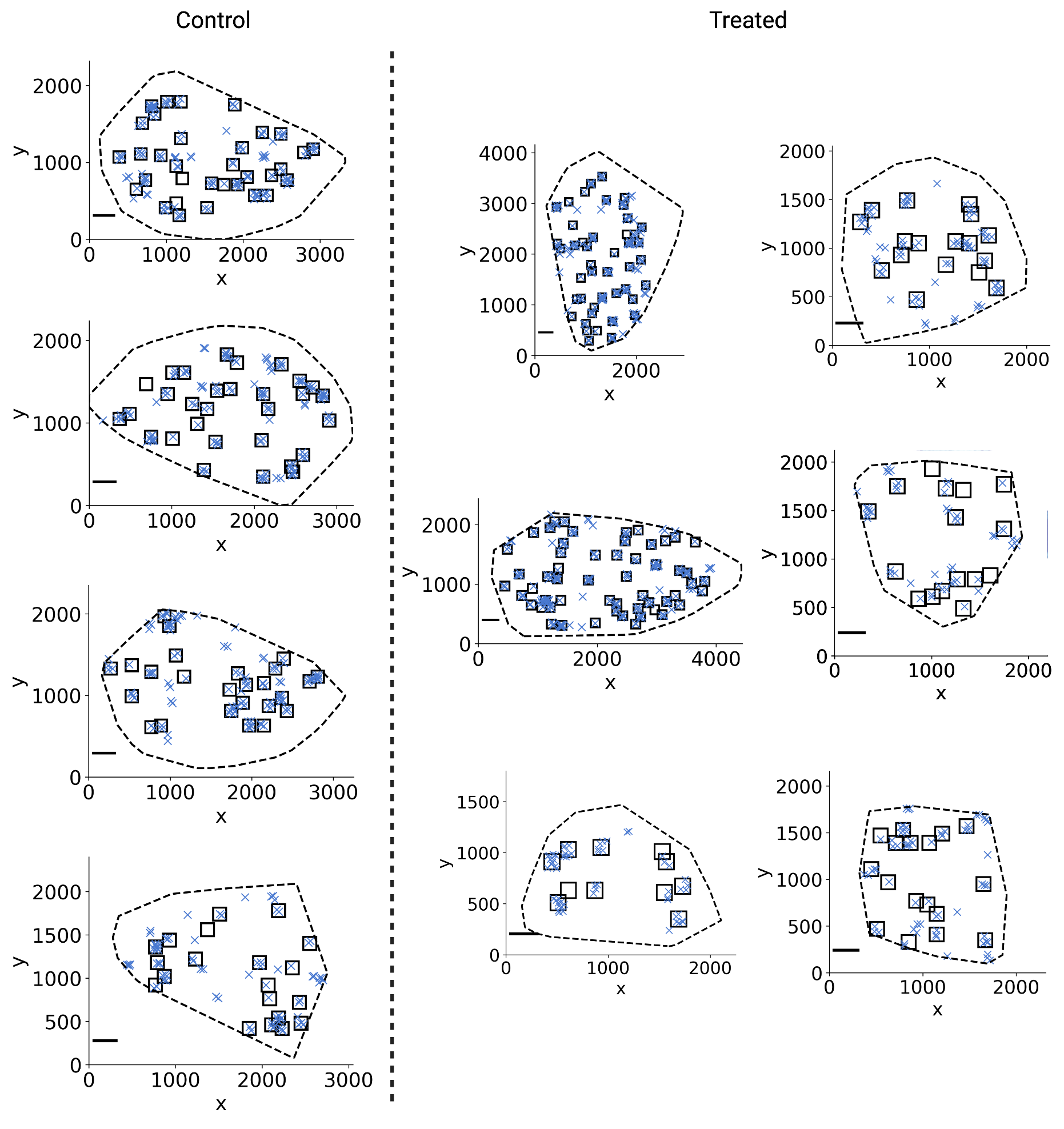}
\caption{TopACT predicted podocyte cells (blue cross) and ground truth glomeruli (black square) for each sample. Note that predicted podocytes colocalize with glomeruli, as expected, validating the use of TopACT on mouse kidney data. Dashed black lines show samples boundaries as in Figure~\ref{fig:masks}. Scale bars = 0.2mm.}
\label{fig:glompreds}
\end{figure}

We then used cell localization pipeline to extract single podocyte cell loci from \methodname{} output. Figure \ref{fig:glompreds} shows that these predicted podocyte cells strongly colocalize with the ground truth glomeruli, validating the use of TopACT on these data.

\begin{figure}
\centering
\includegraphics[width=15.77cm]{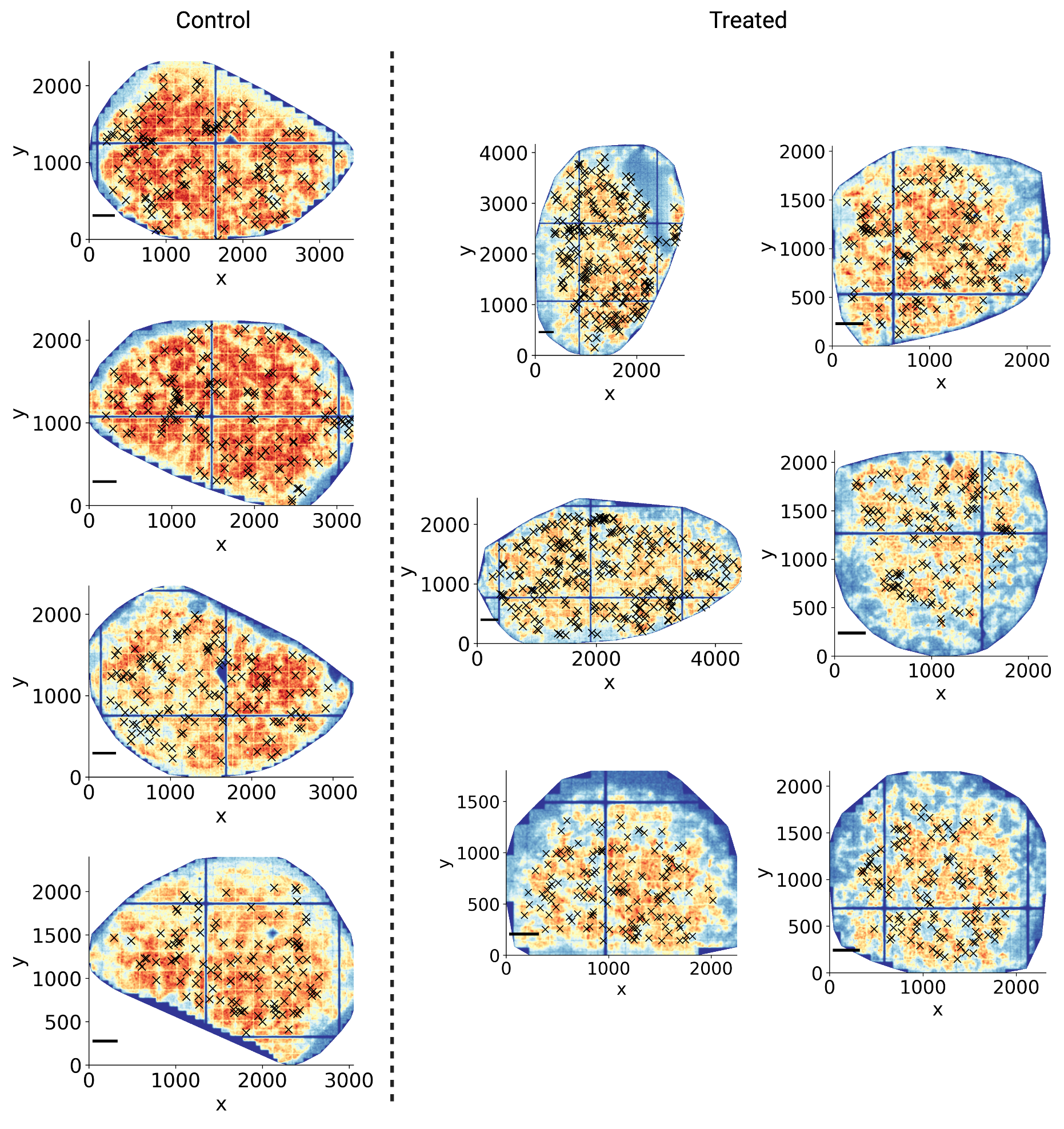}
\caption{TopACT predicted immune cells (black cross) for each sample. Background heatmaps show smoothed transcript count. Scale bars = 0.2mm.}
\label{fig:immunepreds}
\end{figure}

\paragraph{Immune cell predictions} We identified single immune cell loci using our cell localization pipeline. These predictions are shown in Figure \ref{fig:immunepreds}. See the main text (\textit{TopACT pinpoints immune cells in mouse kidney tissue}) for discussion.

\begin{figure}
\centering
\includegraphics[width=15.77cm]{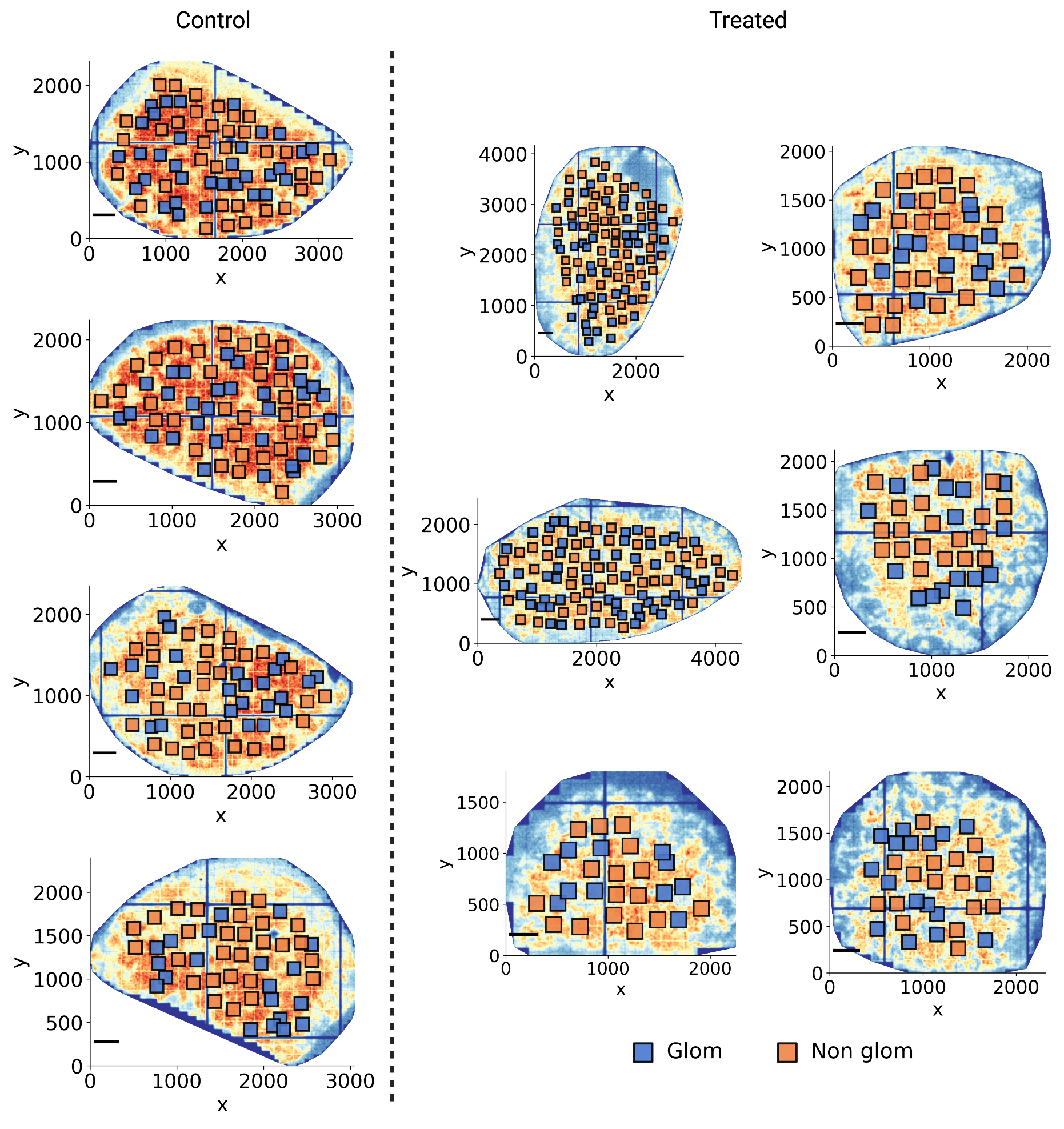}
\caption{Glomerular (blue) and non-glomerular (orange) patches defined on each sample. Each square patch has side length 150 spots (107.25µm). Background heatmaps show smoothed transcript count. Scale bars = 0.2mm.}
\label{fig:patches}
\end{figure}

\paragraph{Generating patches} To normalize for area and to facilitate comparison between glomerular and non-glomerular regions, we split each sample into square patches of side length $150$ spots. Each of these patches is either glomerular (centered on a glomerulus) or non-glomerular (disjoint from any glomerular region).
The glomerular patches were centered on the glomeruli locations extracted from Bin 20 data. To produce non-glomerular patches, we then randomly sampled non-overlapping patches to saturate the remaining area inside these boundaries. We repeated this process several times for each sample, and selected the resulting patch decomposition with the most coverage (i.e.\ the greatest number of patches). We then discarded glomerular patches with less than 90\% overlap with the sample boundaries defined earlier and shown in Figure \ref{fig:masks}. This yielded 269 glomerular patches (108 control, 161 treated) and 310 non-glomerular patches (130 control, 180 treated) (see Figure \ref{fig:patches})

\paragraph{MPH landscapes}
For each patch, we produced a point cloud where points correspond to spots classified as Immune by \methodname{}. For each of these point clouds, multiparameter persistence homology of the Rips-codensity filtration was computed with RIVET \cite{rivet1}, from which MPH landscapes were computed using the code from \cite{vipond2021Multiparameter1}. For the filtration, we used $\rho_5$ for codensity (see Definition \ref{def:codensity}) and set the maximum Rips radius to 100 spots. In RIVET we set the resolution parameter to 30. We then computed average multiparameter persistence landscapes (as in \eqref{eq:mpaverage}) for control and treated glomerular patches. See main text (\textit{MPH characterizes immune cell spatial organization}) for discussion.

\FloatBarrier

\section*{References}

\def\bibfont{\small}

\end{document}